\documentclass[prd,preprint,groupedaddress,onecolumn,showpacs,10pt]{revtex4-1}
\usepackage{hyperref}
\usepackage{graphicx,xcolor}
\usepackage{xtab,afterpage}
\usepackage{lipsum}
\usepackage{bm}
\usepackage{amssymb,amsmath,amsfonts, mathrsfs}

\newcommand{\als}[1]{\begin{align*}#1\end{align*}}
\newcommand{\aln}[1]{\begin{align}#1\end{align}}
\newcommand{\blr}{ {\mathbf r} }

\newcommand{\hatj}{{\hat \jmath}}
\newcommand{\be}{\begin{equation}}
\newcommand{\ee}{\end{equation}}
\newcommand{\ba}{\begin{array}}
\newcommand{\ea}{\end{array}}
\newcommand{\bqa}{\begin{eqnarray}}
\newcommand{\eqa}{\end{eqnarray}}

\begin{document}
\title{On the generalized Friedrichs-Lee model with multiple discrete and continuous states}



\author{Zhiguang Xiao}
\email[]{xiaozg@scu.edu.cn, corresponding author}
\affiliation{College of Physics, Sichuan University, Chengdu  610065, P.~R.~China}

\author{Zhi-Yong Zhou}
\email[]{zhouzhy@seu.edu.cn, corresponding author}
\affiliation{School of Physics, Southeast University, Nanjing 211189,
P.~R.~China}

\begin{abstract}
In this study, we present several improvements of the non-relativistic
Friedrichs-Lee model with multiple discrete and continuous
states and still retain its solvability. Our findings
establish a solid theoretical basis for the exploration of resonance
phenomena in scenarios involving multiple interfering
states across various channels. The scattering amplitudes associated
with the continuum states naturally adhere to coupled-channel
unitarity, rendering this framework particularly valuable for
investigating hadronic resonant states appearing in multiple
coupled channels. Moreover, this generalized framework exhibits
a wide-range applicability, enabling investigations into resonance
phenomena across diverse physical domains, including hadron physics,
nuclear physics, optics, and cold atom physics, among others.
\end{abstract}

\maketitle
\section {Introduction}
Unstable states constitute a ubiquitous phenomenon in contemporary
physics, manifesting across various disciplines such as molecular
physics, nuclear physics and particle physics. In the realm of
hadronic physics, the prevalence of unstable resonances is
particularly notable within the context of strong interactions, where
new resonant states are frequently encountered and documented. These
resonances assume significant significance in unraveling the
fundamental characteristics of hadrons and their interactions,
perpetuating their investigation as a vibrant research area within the
field of particle physics.

To explore the characteristics of unstable states across diverse
branches of physics, several models sharing a similar conceptual
framework have independently emerged. Among these models, the
Friedrichs model stands as a simple non-relativistic Hamiltonian that
couples a bare discrete state to a bare continuous state
~\cite{Friedrichs:1948}. Within this model, the solutions for unstable
generalized eigenstates can be rigorously obtained and expressed in
terms of the bare states. In the realm of quantum field theory, the
Lee model was developed to investigate the properties of field
renormalization~\cite{Lee:1954iq}. This model considers two nucleon
states, denoted as $N$ and $V$, which can be converted to each other
by absorbing or emitting a bosonic $\theta$ particle through the
processes $V\rightleftharpoons N+\theta$.  Analogous models can be
found in various domains, such as the Jaynes-Cummings model in quantum
optics~\cite{JaynesCummings} and the Anderson model in condensed
matter physics~\cite{PhysRevLett.86.2699}.  In this article, we
collectively refer to these models as the Friedrichs-Lee (FL) model,
highlighting their common conceptual foundation. The generalized
eigenstates of the full interacting Hamiltonian within the FL model
can be explicitly determined in terms of the original discrete state
and the continuum states.

The original Friedrichs-Lee model, which involves only one discrete
and one continuous state, is often considered as a toy model due to
its simplicity. It is usually employed to comprehend the properties of bound
states, virtual states and resonant states that appear in the scattering processes.
When the bare discrete state is above the continuum threshold, its
pole position
moves to the second sheet and become a pair of resonance poles. If the
bare discrete state is below the threshold, there would be an
accompanied virtual state pole on the second sheet when the
interaction is turned on. Besides these states generated from the bare
discrete states, there could also be dynamically generated states from
the singularities of the interaction vertices~\cite{Xiao:2016dsx}.
 The mathematical background of describing the unstable states is  the
Rigged Hilbert Space (RHS) quantum
mechanics~\cite{Bohm:1989,Gadella:2004,Prigogine:1991}, rather than  the
conventional Hilbert space. In the RHS quantum mechanics, the
Hamiltonian $H$, as an Hermitian operator, could have generalized complex
eigenvalues and the related eigenstates corresponding to the pole of
the $S$-matrix that lies on the
unphysical sheet of the analytically continued energy plane, commonly referred to as the Gamow states.
The Friedrichs model was also
extended to include more continuous or discrete states and with a more
realistic interaction vertex function. As a result, it finds extensive application in a wide range of realistic scenarios, particularly in the study of hadronic
scattering
processes~\cite{Xiao:2016wbs,Xiao:2016mon,Zhou:2017dwj,Zhou:2017txt,Zhou:2020moj,Zhou:2020vnz}.
Furthermore, coupled-channel models sharing similar spirits have demonstrated success in describing a variety of resonance phenomena in different
physical
systems~\cite{Eichten:1979ms,Tornqvist:1995kr,Kalashnikova:2005ui,Li:2009ad,Ortega:2009hj,Cao:2014qna,Giacosa:2019ldb,Wolkanowski:2015jtc,Wolkanowski:2015lsa,Wang:2018atz,Yang:2021tvc}.
The widespread applicability and efficacy of these models in describing resonance
phenomena render them as powerful tools in studying the properties
of unstable states in different physical contexts.

In the hadron physics, the usual effective field theory calculation of
the scattering amplitude encounter challenges pertaining to unitarity
and analyticity. The perturbative $S$ matrix generally fails to
generate bound states or resonance poles on the analytically continued
Riemann surface of the energy plane. To address this,
various unitarization methods are used, such as the $K$-matrix
method. The typical $K$-matrix parameterization of the $S$-matrix like
$S=\frac {1-iK}{1+iK}$ lacks a dynamical origin and enforces unitarity
by hand. However, this parametrization does not guarantee the absence of
unphysical spurious poles, including those situated in
the complex energy plane of the first Riemann sheet, which
violates causality. In contrast, the Friedrichs-Lee model achieves
unitarity as a consequence of its dynamics, and the Hermitian property
of the Hamiltonian ensures the absence of spurious poles in  the first
Riemann sheet. These are the immediate advantages of these kinds of models over the $K$-matrix parameterization.

While there have been notable achievements in the application of such
models, certain aspects still call for further improvement.
From a quantum field theory perspective, the previous model only considered
the contribution of intermediate $s$-channel  discrete state to the
amplitude.  However, there are other types of interactions that are not
included. The first one comes from the crossed channels in the
two-to-two scattering amplitude, where the intermediate particle can also appear
as the $t$- or $u$-channel propagators.  The second one involves the contact
interactions, such as the four-point vertex.  Upon performing a  partial wave
projection, both of these interactions can be represented by
continuum-continuum interactions. These interactions introduce a mild
background to the final experimental observation, potentially
interfering with the $s$-channel resonances and modifying the lineshape.
It is crucial to include these background contributions in the analysis of the experimental data while preserving
analyticity and unitarity.  The commonly used Breit-Wigner parametrization to
parameterize the $t$- or $u$-channel resonance and a polynomial to
parameterize the background would violate the unitarity. A na\"ive $K$-matrix unitarization may introduce
unexpected spurious poles in the $S$-matrix. Thus, to
incorporate the continuum-continuum interactions into the
Friedrichs-Lee like models could overcome these problems.  However, a
general continuum-continuum interaction renders the model no longer
solvable.  In refs.~\cite{Xiao:2016mon,Sekihara:2014kya}, a particular
form of separable interaction involving the continuum states is
introduced, where  the interaction vertex function between the
discrete states and the continuum  also appears as the factors of the
separable interaction between two continuum states. The Hamiltonian
is
\begin{align}
H=&\sum_{i=1}^D M_i|i\rangle\langle
i|+\sum_{i=1}^N \int_{a_i}^\infty \mathrm d \omega
\,\omega|\omega;i\rangle\langle \omega;i|\nonumber
\\ &+\sum_{i,j=1}^C v_{ij}\Big(\int_{a_i}^\infty\mathrm d \omega
f_i(\omega)|\omega;i\rangle\Big)\Big(\int_{a_j}^\infty\mathrm d \omega
f^*_j(\omega)\langle \omega;j|\Big)\nonumber
\\ &+\sum_{j=1}^D\sum_{i=1}^C \left[ u_{ji}^*|j\rangle\Big(\int_{a_i}^\infty\mathrm d \omega
f^*_i(\omega)\langle \omega;i|\Big)+ u_{ji}\Big(\int_{a_i}^\infty\mathrm d \omega
f_i(\omega)|\omega;i\rangle \Big)
\langle j|
\right],
\end{align}
where the form factor $f_i(\omega)$  is associated with the
$i$-th continuum
state $|\omega;i\rangle$, both for its the interaction with different
discrete states and for its interaction with other continuum states.
There are two aspects which could be
improved for this model. First, the interaction between the discrete
states $|j\rangle$ and the continuum could be extended to  a general
function $f_{ij}(\omega)$, for a more realistic description of the
strong interaction in the real world.  Using the quark pair
creation~(QPC) model as an example, the interaction between a meson
and their decay products is expressed as a complicated integration
between the wave function for the three states and the pair production
vertex~\cite{Micu:1968mk,Zhou:2017txt}. Thus the form of the
interaction function depends both on the discrete state and the
continuum. Secondly, the interaction
between the continuum states does not need to be factorized using the
same factors as the interaction between the discrete state and the
continuum. In this paper, we will demonstrate that after factorizing the
continuum-continuum interaction independent of the discrete-continuum
interaction, the model remains exactly solvable. In principle, the extra
continuum-continuum interaction should be the residue interaction
 after subtracting the $s$-channel intermediate discrete
state contribution, which could have no relation with the
discrete-continuum interaction. Whether this interaction can be
expressed as a separable potential remains an open question.
There are aleady some physical applications of the separable
potentials in discussing real world problems, for example, describing
the interaction between the open-flavor channels and the hidden-flavor
channels in momentum space~\cite{Guo:2016bjq}.
Our formalism differs from this implementation by two key
advances: First, we parameterize all continuum-continuum couplings via
separable potentials, withou the distinction between open-flavor and
hidden -flavor channels. This enables a more general description of
coupled-channel systems. Secondly, by projecting potentials onto
angular momentum eigenstates through spherical harmonic expansion, the
three-dimensional momentum integration reduces to a one-dimensional
radial integral. This systematically eliminate angular variables,
significantly simplifying both numerical implementation and analytical
discussion of mementum dynamics.  Moreover, in general, a
square-integrable interaction potential between
continuum states could be expanded using  a series of general separable
basis. In fact, one can also expand both the
discrete-continuum interaction vertex and continuum-continuum
interaction vertices using the same function basis.
Thus, the study of such separable potentials may have broader physical
applications.  In this paper, our focus lies on these two kinds of
improvements: the
incorporation of a more general
discrete-continuum interaction and various separable
continuum-continuum interactions among multiple bare discrete and continuum
states in the FL model.  By rigorously solving the
eigenstates for the Hamiltonian, we obtain the ``in'' and ``out''
states, the scattering S-matrix, discrete state solution, and other
mathematical physics properties. Our aim is to establish a solid
foundation for the further phenomenological applications of the FL
model by  including these additional physical features.

We organize the paper as follows: In Section \ref{sect:dis-cont}, the solution of the
FL model with more general interactions between discrete states and
continuum states is derived. Section \ref{sect:sep-cont} discusses the case with extra separable
continuum-continuum interactions.  Section \ref{sect:Approx-Sep} is devoted to studying the
case when the interaction potential between continuum states could be
approximated by a sum of separable potentials and consider the
cases when both the continuum-continuum potential and continuum-discrete
potentials are approximated by a truncated series. In section \ref{sect:discuss}, as an
application, we consider some simple examples and discuss the behavior
of the discrete states after turning on various interaction. Section
\ref{sect:conclude} is the conclusion.

\section{The extended Friedrichs-Lee model with multiple discrete
states and continuum states\label{sect:dis-cont}}

First, we are going to consider a system with $D$ kinds of
discrete states and $C$ kinds of continuum states, where $C$ and $D$
denote the numbers of the continuum states and the discrete states
respectively. If there is no interaction, the mass of the $j$-th discrete state
$|j\rangle$ is $M_j$, while the energy spectrum of  the $n$-th continuum
state ranges in $[a_n,\infty)$ with the threshold energy $a_n$.
The interaction between the $j$-th discrete state and the $n$-th
continuum state can be generally represented by a coupling
function $f_{jn}(\omega)$. The full Hamiltonian can be expressed as
\begin{align}
H=&H_0+H_I,
\end{align}
where the free Hamiltonian $H_0$  could be written down explicitly as
\begin{align}
H_0=&\sum_{i=1}^D M_i|i\rangle\langle
i|+\sum_{n=1}^C \int_{a_n}^\infty \mathrm d \omega
\,\omega|\omega;n\rangle\langle \omega;n|,
\end{align}
and the interaction part $H_I$ reads
\begin{align}
H_I=&\sum_{j=1}^D\sum_{n=1}^C \left[ |j\rangle\Big(\int_{a_n}^\infty\mathrm d \omega
f^*_{jn}(\omega)\langle \omega;n|\Big)+ \Big(\int_{a_n}^\infty\mathrm d \omega
f_{jn}(\omega)|\omega;n\rangle \Big)
\langle j|
\right].
\end{align}
The free eigenstates are supposed to be orthogonal to each other and
the normalization conditions satisfy $\langle i|j\rangle=\delta_{ij}$,
$\langle i|\omega;n\rangle=0$ and $\langle
\omega;n|\omega';n'\rangle=\delta(\omega-\omega')\delta_{nn'}$. For
simplicity, we first suppose that there is no degenerate threshold and no
degenerate discrete states.  In fact, if there are degenerate states
with the same threshold and the same
interactions with the other states, the corresponding solutions will
also be  degenerate with the same expression after the
interactions are turned on, and we will take them as one state
with degenerate degrees of freedom just like different magnetic
quantum numbers when there is no magnetic field.
If the states with degenerate
threshold take part in different interactions, the following
discussion will not be modified too much.  We will come back to this case
later.

The general solution for the energy eigenvalue problem
$H|\Psi(E)\rangle=E|\Psi(E)\rangle$ can be represented as a linear
combination of the discrete states and the continuum states,
\begin{align}\label{eq:expand-1}
|\Psi(E)\rangle=\sum_{i=1}^D
\alpha_i(E)|i\rangle+\sum_{n=1}^C\int_{a_n} \mathrm d\omega
\psi_n(E,\omega)|\omega;n\rangle,
\end{align}
where the $\alpha_i(E)$ and $\psi_n(E,\omega)$ functions are defined as the coefficient functions of the discrete states and the continuum states respectively.
By substituting this ansatz into the eigenvalue equation, and carefully examining  the coefficients preceding the discrete states and the continuum states,
we can derive two distinct sets of equations,
\begin{align}
&(M_j-E)\alpha_j(E)+\sum_{n=1}^C \int_{a_n}^\infty
\mathrm d \omega f^*_{jn}(\omega)\psi_n(E,\omega)=0\,,\quad
\mathrm{for } \ j=1,\dots,D
\label{eq:eigen-s-eqs-1}\\
&\sum_{j=1}^D\alpha_j
(E)f_{jn}(\omega)+(\omega-E)\psi_n(E,\omega)=0\,, \quad \mathrm{for} \
n=1,\dots C,\ \mathrm{and}\  \omega>a_n.
\label{eq:eigen-s-eqs-2}
\end{align}
An important observation to make is that the formula exhibits a
nontrivial complexity, which does not appear in the single-channel
scenario. Specifically, for a given energy range $a_l < \omega < a_{l+1}$, there are only $l$ equations present in Eqs.~(\ref{eq:eigen-s-eqs-2}).

Consequently, the eigenvalue problem yields both continuum solutions and discrete solutions. These solutions correspond to different regimes of the spectrum, which will be addressed carefully in the following.

\begin{enumerate}
  \item  \textbf{The continuum state solutions}

   When the energy $E$ is above the highest threshold, that means,  $E>a_C$,  there will be $C$ continuum states when the interactions are turned on, so the $m$-th continuum solution will be
\begin{align}
|\Psi_m(E)\rangle=\sum_{i=1}^D
\alpha_{mi}(E)|i\rangle+\sum_{n=1}^C\int_{a_n} \mathrm d\omega
\psi_{mn}(E,\omega)|\omega;n\rangle, \quad m=1,2,\dots,C\,.
\label{eq:expand-Cont}
\end{align}
However, when the energy $E$ is lower than the highest threshold, e.g., $E\in[a_l,a_{l+1})$, $l<C$, there will be $l$ degenerate continuum
eigenstates, $m=1,2\dots,l$, and the other states are not well-defined
below their thresholds and are set to $\mathbf 0$. To remove the ambiguity of the degenerate
states, it is required that when the interaction is
turned off, i.e. $f_{jm}(\omega)\to 0$,
$|\Psi_m\rangle$ tends to the free continuum state  $|E;m\rangle$.
We expect that, when the eigenvalue $E\in[a_1,a_2]$, we can
solve $\alpha_{1,i}$ and $\psi_{1,i}$ in $|\Psi_1(E)\rangle$, and then
analytically extend these parameters
to $E\in[a_2,a_3]$ to solve $|\Psi_2(E)\rangle$, and so on. In this
way the eigenfunctions can be uniquely determined. From Eq.~(\ref
{eq:eigen-s-eqs-1},\ref{eq:eigen-s-eqs-2}) in terms of the coefficients in
Eq.~(\ref{eq:expand-Cont}), the coefficient function
$\psi_{mn}(E,\omega)$ before the continuum state  in different energy regions could be expressed as
\begin{align}
(\mathrm{ for}\ n\le l) \quad
\psi_{mn}(E,\omega)=&\gamma_n\delta_{mn}\delta(\omega-E)+\frac
1{E-\omega\pm i 0}\sum_{j=1}^D \alpha_{mj}(E)f_{jn}(\omega)\,,\nonumber
\\
(\mathrm{ for}\ n> l) \quad \psi_{mn}(E,\omega)=&\frac
1{E-\omega\pm i 0}\sum_{j=1}^D \alpha_{mj}(E)f_{jn}(\omega)\,.\nonumber
\end{align}
This equation could be concisely written down in one equation by using the Heaviside step
function $\Theta(x)$,
\begin{align}\label{eq:coeffunc1}
\psi^\pm_{mn}(E,\omega)=&\gamma_n\delta_{mn}\delta(\omega-E)\Theta(E-a_n)+\frac
1{E-\omega\pm i 0}\sum_{j=1}^D f_{jn}(\omega)\alpha_{mj}(E)\,.
\end{align}
Notice  that $\psi^{\pm}_{mn}$ is a generalized function, and in order to distinguish between different integral contours, we have included $\pm i0$ in the denominator of the integral in
Eq.(\ref{eq:expand-Cont}). The $\psi^+$ state corresponds to the coefficient for
the in-state while
$\psi^-$ corresponds to those of the out-state.
For the convenience of the future discussions, we will omit the
superscripts $\pm$ in the notations. It should be understood that
the appropriate superscript can be easily inferred based on the
context. In the cases where there is a need to explicitly indicate the
in-state or out-state, we will make use of the superscript
accordingly.

Inserting this equation back into Eq.(\ref{eq:eigen-s-eqs-1}), we can
obtain the equations for the coefficient functions $\alpha_{mk}(E)$
 for $m=1,2,\dots l$
\begin{align}\label{eq:coeffunc2}
&-\sum_{k=1}^D\alpha_{mk}(E)\Big[\delta_{kj}(E-M_j)-\sum_{n=1}^C\int_{a_n}^\infty\mathrm
d\omega\,
\frac{f_{kn}(\omega)f^*_{jn}(\omega)}{E-\omega\pm i0}\Big]+
\sum_{n=1}^C\gamma_m(E)\delta_{mn}f^*_{jn}(E)=0.
\end{align}

With many different discrete states and continuum ones involved in,
the representation becomes much more complex than the simplest
version. In fact, the formula and the derivation procedure could be
simplified by introducing the matrix form. In the following, the
matrices are represented in bold faces and the dot symbol ``$\cdot$"
represents the matrix product, and the matrix element is expressed in
the form like $(\bm\eta)_{ij}$. For example, Eq.~(\ref{eq:coeffunc2}) could be written down in the matrix form as
\begin{align}
- \bm \alpha^\pm(E)\cdot \bm \eta_\pm(E)+\bm
\gamma(E)\cdot
 \mathbf f^\dagger(E) =0\,,\nonumber
\end{align}
where $\bm \alpha$ and $\mathbf f$ are the $C\times D$ and $D\times C$ matrices for the
coefficients
$\alpha_{mk}$ 
and $f_{jm}$, respectively.
The matrix $\bm\gamma$ is defined as a  diagonal matrix of dimension $C\times C$
\bqa
(\bm \gamma)_{mn}(E)=\gamma_n\delta_{mn}\Theta(E-a_n),\nonumber
\eqa
whose diagonal elements $\gamma_n$ could be different in principle for
different $n$ and the values could be determined by  the normalization
conditions.  The $\bm\eta_\pm$ matrix, the inverse of resolvent
function matrix, is of dimension $D\times D$ and every matrix element
reads
\aln{(\bm \eta_\pm)_{kj}(E) =&(E-M_j)\delta_{kj}-\sum_{n=1}^C\int_{a_n}^\infty\mathrm
d\omega\, \frac{f_{kn}(\omega)f^*_{jn}(\omega)}{E-\omega\pm i0}.
\label{eq:eta-matrix}}

In general,  the
determinant of $\eta$ matrix does not vanish  for $a_l<E<a_{l+1}$, and the matrix
$\bm\alpha^\pm$ can be represented as
\aln{\bm \alpha^\pm(E)=
\bm \gamma(E)\cdot
\mathbf f^\dagger(E)\cdot\bm \eta_{\pm}^{-1}(E). \nonumber}
Inserting this result into Eq.~(\ref{eq:coeffunc1}), the coefficient
functions $\psi_{mn}$ before the continuum states can be obtained  in matrix representation
\als{
\bm \psi^\pm(E,\omega)=&\bm\gamma\delta(\omega-E)+\frac
1{E-\omega\pm i0}\bm \gamma(E)\cdot
\mathbf f^\dagger(E)\cdot\bm \eta_\pm^{-1}(E)\cdot \mathbf f(\omega).
}

The solution of the continuum eigenstate can then be expressed as
\begin{align}
|\Psi^\pm_m(E)\rangle=&\sum_{i=1}^D
\alpha^\pm_{mi}(E)|i\rangle+\sum_{n=1}^C\int \mathrm d\omega
\psi^\pm_{mn}(E,\omega)|\omega;n\rangle\nonumber
\\=&\gamma_m\Theta(E-a_m)|E,m\rangle+\sum_{k=1}^D\big(
\bm \gamma(E)\cdot
\mathbf f^\dagger(E)\cdot\bm
\eta_\pm^{-1}(E)\big)_{mk}\Big(|k\rangle+\sum_{n=1}^C\int_{a_n}d\omega\frac
{f_{kn}(\omega)}{E-\omega\pm i0}|\omega;n\rangle\Big)\,.
\label{eq:Cont-m}
\end{align}
Notice that in the energy region $a_l<E<a_{l+1}$, the wave function
$|\Psi^\pm_m\rangle$ for $m>l$ should vanish. Another required condition
is that, when the coupling function $f_{jn}$ vanishes,
$|\Psi^\pm_m(E)\rangle $ tends to $|E,m\rangle$. Therefore, the
coefficient $\gamma_m$ is determined to be 1.
It can be checked that  the normalization satisfies $\langle
\Psi^\pm_m(E)|\Psi^\pm_n(E')\rangle=\delta(E-E')\delta_{mn}$.
Actually, in the point view of the scattering theory, $|\Psi^+\rangle$ is the ``in"
state and $|\Psi^-\rangle $ is the ``out"
state, so the $S$ matrix can be obtained by inner product of the  ``in" state and the ``out" state as
\aln{
\langle
\Psi^-_m(E)|\Psi^+_n(E')\rangle&=\gamma_m^*\gamma_n\delta(E-E')
-2\pi i\delta(E-E')\big(\bm\gamma(E')\cdot 
\mathbf
f^\dagger(E')\cdot\bm
\eta_+^{-1\dagger}(E)\cdot
\mathbf f(E)
\cdot\bm \gamma^\dagger(E)
\big)_{nm}
\nonumber\\
&=\delta(E-E')\big[
\bm \gamma\cdot\big(1
-2\pi i \mathbf
f^T(E)\cdot\bm
\eta_+^{-1}(E)\cdot
\mathbf f^*(E)
\big)\cdot \bm \gamma\big]_{nm}.}
The $\bm\eta_\pm(E)$ function can be analytically extended  to the
complex $E$ plane with $\eta_+(E)$ and $\eta_-(E)$ coinciding with
$\bm \eta(E)$ on the upper edge and lower edge of the real axis
 above the thresholds, respectively. We can also define the
analytically continued $S$ matrix
\bqa \label{Smatrix1} \mathbf S=1
-2\pi i \mathbf
f^T(E)\cdot\bm
\eta^{-1}(E)\cdot
\mathbf f^*(E),
\eqa
where $E$ is analytically continued to the complex energy plane and
only when $E$ is real and on the upper edge of the cut above the
lowest threshold $a_1$ is the $S$ matrix the physical one.
{ Given the presence of $C$ continuous states with distinct thresholds,
it is a general result that there exist $2^C$ different Riemann sheets
for the analytically continued $S$ matrix. Since only the Riemann
sheets nearest to the physical region affect the physical $S$-matrix
the most, we label the $m$-th sheet as the Riemann sheet continued
from the physical region $(a_m,a_{m+1})$, where the first sheet where
the physical $S$ matrix resides is called physical sheet by
convention.}

It is worth pointing out that the formula of scattering matrix
Eq.~(\ref{Smatrix1}) has important phenomenological applications.  For
example, in studying the particle-particle scattering processes, the
two particles that collides or those final states~(usually called as
channels in the scattering experiments) form continuum states, while
the intermediate resonance states are regarded as the discrete states.
The $(n,m)$-th element of scattering matrix in Eq.~(\ref{Smatrix1})
could describe the scattering amplitudes from the  channel of $n$-th
continuum state to the $m$-th  channel. The coupled-channel unitarity
is naturally satisfied among all the related scattering amplitudes due
to the obvious relation  $\mathbf S\mathbf S^\dag= I$. Furthermore, once
the coupling function between the discrete and continuum state is
reliably described by some dynamical models, the physical observables,
such as the cross sections, could be predicted or
calculated~\cite{Zhou:2023yjv}.

  \item \textbf{The discrete state solutions:}

In Eqs.~(\ref{eq:eigen-s-eqs-1}, \ref{eq:eigen-s-eqs-2}), if the eigenvalue $E\notin [a_n,\infty)$ for $n=1,\dots, C$,
there is no need to introduce the $\pm i 0$ in the denominator of the integrand and we have
\begin{align}
\psi_n(E,\omega)=&\frac 1{E-\omega}\sum_j f_{jn}(\omega)\alpha_j(E)\,, \quad (\mathrm{for} \ n=1,\dots C)
\label{eq:cont-eq-2}\\
(\bm\alpha(E)\cdot \bm \eta(E))_j=&\sum_{k=1}^D\alpha_k(E)\Big[\delta_{kj}(M_j-E)-\sum_{m=1}^C\int_{a_m}^\infty\mathrm
d\omega\, \frac{f_{km}(\omega)f^*_{jm}(\omega)}{\omega-E}\Big]=0\,,\quad
(\mathrm{for } \ j=1,\dots,D)\,.
\label{eq:discret-eq-2}
\end{align}

In order to obtain nonzero solutions of $\alpha_k(E)$, it is necessary to satisfy the condition $\det
\bm\eta(E)=0$. This condition implies that there may exist  discrete energy solutions for this equation, which in general correspond to the poles of the
$S$-matrix elements.
If there exist  solutions on the first sheet, they must reside on the real
axis below the lowest threshold since the eigenvalue of a hermition
Hamiltonian for a normalizable eigenstate should be real. Additionally,  solutions can also be found on the
unphysical sheets, which may corresponds to complex conjugate resonance
poles on the complex energy plane or to virtual state poles located on the real axis below the lowest
threshold. There would
be at least $D$ discrete solutions which tend to the bare discrete states, i.e.  $\alpha^{(l)}_k\to
\delta_{kl}$ and $E\to M_l$ for $l=1,2,\dots, D$, as all the coupling function $f_{ln}\to 0$. Furthermore, it is possible for other dynamically generated states that do not go to
the bare states when the interactions are switched off. In general, the
solutions does not exhibit degeneracy, indicating that the poles for $S$
matrix are just simple poles. If the degenerate solutions occur for
$\det \bm\eta(E)=0$, it implies that  two or more poles may coincide
and form a higher order pole. This situation is considered to be
accidental and only occurs for some special coupling functions. For
the purposes of our discussion, we will not consider this
special case and assume that the solutions are non-degenerate. Then
for each energy solution $E_i$, we can also find the eigenvector
$\alpha^{(i)}_k(E_i)$ and $\psi^{(i)}_n(E_i,\omega)$, and the wave function of discrete state is expressed as
\begin{align}
|\Psi^{(i)}(E_i)\rangle=\sum_{j=1}^D
\alpha^{(i)}_j(E_i)\Big(|j\rangle+\sum_{n=1}^C\int_{a_n} \mathrm d\omega
\frac{f_{jn}(\omega)}{E_i-\omega}|\omega;n\rangle\Big).
\label{eq:Discrete-Solu}
\end{align}
When $E_i$ lies on the real axis of the first Riemann Sheet below the
lowest threshold, this wave function corresponds to a bound state. In this case, the
integrals in $\eta(E_i)$ and $\alpha_k^{(i)}(E_i)$ are real. The
normalization for this state is well-defined and the
$\alpha_k^{(i)}(E_i)$ can be chosen such that
\aln{ 1=&\sum_{jk}^{D}
\alpha_k^{(i)}(E_i)\Big(\delta_{jk}+\sum_{m=1}^C\int_{a_m}d\omega\frac{f_{km}(\omega)f^*_{jm}(\omega)}{(E_i-\omega)^2}\Big)\alpha_j^{(i)*}(E_i)\nonumber
\\=&\sum_{jk}^D\alpha^{(i)}_{k}(E_i)\eta_{kj}'(E_i)\alpha^{(i)*}_{j}(E_i)\,,
\label{eq:Norm-i}
}
with $\eta'_{kj}(E)$ being the derivative of $\eta_{kj}(E)$ w.r.t.
$E$. Actually, this equation has a probabilistic explanation. The first term on the right-hand side of the equal sign represents the probability of finding the bare discrete states in the bound state, while the second one represents those of finding the bare
continuum states in it. If we define
\aln{Z^{(i)}_k=|\alpha_k^{(i)}(E_i)|^2,\quad
X^{(i)}_m=\sum_{k,j=1}^D\alpha_k^{(i)}(E_i)\Big(\int_{a_m}d\omega\frac{f_{km}(\omega)f^*_{jm}(\omega)}{(E_i-\omega)^2}\Big)\alpha_j^{(i)*}(E_i),}
then,  $Z^{(i)}\equiv \sum_k Z^{(i)}_k$ is called the elementariness
and   $X^{(i)}\equiv \sum_{m=1}^CX^{(i)}_m$ is called the compositeness for the bound
state.
When the solution $E_i$ resides on the unphysical sheet, it is necessary to deform the integral
contour to bypass the pole position in different
integrals, as illustrated in Fig.~\ref{fig:contour}. For resonance poles on the
$m$-th sheet, the integral contour for the first $m$-th integral should
be deformed accordingly, following the contour shown  in  Fig.~\ref{fig:contour}.
In such cases, the usual definition of the normalization may not be well-defined, as the integral contour for the pole and its conjugate pole are not consistent. Therefore, it becomes necessary to define the  normalization through the inner product of the
state and its conjugate state which corresponds to the conjugate pole.
The resulting normalization is similar to Eq. (\ref{eq:Norm-i}) with $E_i$ replaced  by the pole position on the unphysical sheet and the
integral contour suitably deformed. However, it is important to note that the probabilistic interpretation of each term in the sum will no longer hold, as the terms may not be real
or positive for poles on the unphysical plane.
\end{enumerate}
\begin{figure}
\includegraphics[height=2cm]{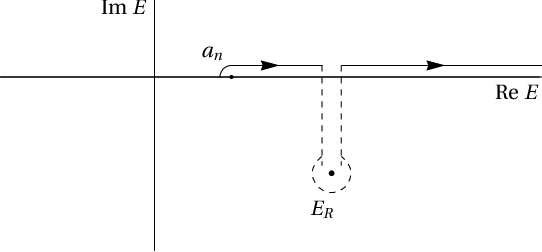}
\caption{ The integral contour for the resonance
solution.\label{fig:contour}}
\end{figure}

Now we come to the case with degenerate threshold. If there are
different continuum states with the same threshold $a_n$, with
degeneracy $h_n$, we need to add
another label $\kappa$ to the continuum to denote the different continuum
states sharing the same threshold, $|\omega,n\kappa\rangle$.
Thus all the indices in the equations labelling the continuum states
would include the additional indices $\kappa$ to label the degenerate
states, for example,
$f_{in}(\omega)$, $\alpha_{ni}$, $\gamma_m$, $\psi_{mn}$, $\Psi_m$  become
$f_{i,n\kappa}$, $\alpha_{n\kappa, i}$, $\gamma_{m\kappa}$, $\psi_{m\kappa,n\kappa'}$.
There will be $\mathfrak h=\sum_{n=1}^C h_n$ continuum
states.
The sum over the continuum states also needs to sum over the $\kappa$.
The matrix $\bm\gamma$ is defined as
$\gamma_{m\kappa,n\kappa'}(E)=\delta_{mn}\delta_{\kappa,\kappa'}\Theta(E-a_n)$.
$\mathbf f$ matrix becomes a $D\times \mathfrak h$ matrix and $\bm
\eta$ is still a $D\times D$ matrix. With all these changes, the
previous discussion and equations can be smoothly used in this case.

\section{Including separable continuum-continuum
interactions \label{sect:sep-cont}}
 In the previous case, we considered a scenario where a bare
continuum state is only coupled to the bare discrete states but not to
the other continuum states. However, when the direct interactions between
continuum states become significant, it is more appropriate to include
the corresponding term in the interaction Hamiltonian $H_I$.
Analytically solving the Hamiltonian with a general
continuum-continuum interaction is generally not feasible. Therefore,
in this section, we will focus on the case with a separable
interaction, which still allows for an exactly solvable solution.

The Hamiltonian, including $D$ discrete states and $C$ continua with
factorizable self-interacting contact terms, can be expressed as
\begin{align}
H=&\sum_{i=1}^D M_i|i\rangle\langle
i|+\sum_{n=1}^C \int_{a_n}^\infty \mathrm d \omega
\,\omega|\omega;n\rangle\langle \omega;n|\nonumber
\\ &+\sum_{m,n=1}^C v_{mn}\Big(\int_{a_m}^\infty\mathrm d \omega
 g_m(\omega)|\omega;m\rangle\Big)\Big(\int_{a_n}^\infty\mathrm d \omega
 g^*_n(\omega)\langle \omega;n|\Big)\nonumber
\\ &+\sum_{j=1}^D\sum_{n=1}^C \left[ |j\rangle\Big(\int_{a_n}^\infty\mathrm d \omega
f^*_{jn}(\omega)\langle \omega;n|\Big)+ \Big(\int_{a_n}^\infty\mathrm d \omega
f_{jn}(\omega)|\omega;n\rangle \Big)
\langle j|
\right].
\label{eq:H-separable}
\end{align}
In this case,  the new coupling constants $v_{mn}$ between two
continuum states have been incorporated in the interaction terms, and
$v_{mn}=v_{nm}^*$ is satisfied to meet the Hermiticity requirement. The
form-factor functions $g_n(\omega)$ are involved in the interaction
between two continuum states, and $f_{jn}(\omega)$ represents the  interaction vertex between the $j$-th discrete
state and the $n$-th continuum state. We suppose $f_{jn}$ matrix is of
full rank, otherwise one can always find a decoupled state by linear
combination of the discrete states or the continuum states.
For the sake of simplicity, we first assume that the coupling
constant matrix $v_{mn}$ is non-degenerate and will come back
to the degenerate $v_{mn}$ case later.

Similar to the previous case, we are also going to solve the
Hamiltonian eigenfunction $H|\Psi(E)\rangle=E|\Psi(E)\rangle$. The
eigenstate of the Hamiltonian with eigenvalue $E$ can be  be
expanded in terms of the discrete
states and the continuum states as
\begin{align}
|\Psi(E)\rangle=\sum_{i=1}^D
\alpha_i(E)|i\rangle+\sum_{n=1}^C\int_{a_n} \mathrm d\omega
\psi_n(E,\omega)|\omega;n\rangle.
\label{eq:expand-2}
\end{align}
Inserting this ansatz into the eigenvalue equation and projecting to
the discrete eigenstates or the continuum ones, one can find two sets
of equations
\begin{align}
& (M_j-E)\alpha_j(E)+A_j(E)=0,\quad
j=1,\dots,D
\label{eq:eigen-s-eq-1}\\
&\sum_{j=1}^D\alpha_j
(E)f_{jn}(\omega)+(\omega-E)\psi_n(E,\omega)+\sum_{m=1}^Cv_{nm}B_m(E)g_n(\omega)=0,
\quad n=1,\dots,C, \quad \mathrm{and\ }\omega>a_n\label{eq:eigen-s-eq-2}
\end{align}
where,   two new integration
functions $A_j(E)$ and $B_n(E)$ have been defined as
\aln{A_j(E)\equiv\sum_{n=1}^C\int_{a_n}^\infty\mathrm
d\omega\, f^*_{jn}(\omega)\psi_n(E,\omega), \quad B_n(E)\equiv\int_{a_n}^\infty\mathrm
d\omega\, g^*_n(\omega)\psi_n(E,\omega).\label{eq:def-A-B}}
Since we have assumed that the continuum-continuum coupling constants $v_{mn}$s are not degenerate, there are $C$ independent $B_n(E)$
functions. On the contrary, if $v_{mn}$ matrix is degenerate, the only change is that there
will be fewer $g_n$ and $B_n$ functions.

Similarly as discussed previously,  if the eigenvalue $E\in [a_l,a_{l+1})$ for $l< C$, there should be $l$ continuum solutions $|\Psi_m(E)\rangle$, $m=1,2,\dots, l$ with the same eigenvalue $E$.
 As the interactions are gradually deactivated, it is required that
these continuum solutions tends to well-defined states $
|E,m\rangle$. This ensures that in the absence of interactions, the
continuum solutions can be uniquely determined as the continuum states
$ |E,m\rangle$, thus eliminating any ambiguity in their
characterization. This requirement guarantees a smooth transition from
the interacting system to the non-interacting system.

Under these specific conditions, the $l$
contiuum state solutions for
$a_l<E<a_{l+1}$  coincide with the first
$l$ states for $E>a_C$. Consequently, it is sufficient to solve for the solutions when $E>a_C$, and then  the first $l$ solutions can be obtained for the range $E<a_l$.
For each continuum state $|\Psi_m(E)\rangle$ with $E>a_C$, the corresponding coefficients are denoted as $\alpha_{jm}$,
$\psi_{nm}(E,\omega)$, $A_{jm}$, and $B_{nm}$ as in eqs.~(\ref{eq:expand-2}), (\ref{eq:eigen-s-eq-1}), and (\ref{eq:eigen-s-eq-2}). Then,  one could obtain
\begin{align}
&\alpha^\pm_{jm}(E)=\frac1{E-M_j}A^\pm_{jm}(E),
\label{eq:discret-eq-cont}
\\
&\psi^\pm_{nm}(E,\omega)=\delta_{nm}\delta(\omega-E)+\frac
1{E-\omega\pm i0}\Big(\sum_{j=1}^D \frac{A^\pm_{jm}(E)
f_{jn}(\omega)}{E-M_j}+\sum_{n'=1}^C v_{nn'}B^\pm_{n'm}(E)g_n(\omega)\Big)
\,.
\label{eq:cont-eq-cont}
\end{align}

The procedure of solving the equation Eq.~(\ref{eq:cont-eq-cont}) is
straightforward but intricate.  The strategy is to apply the
operations $\sum_n\int_{a_n} d\omega f^*_{jn}(E)\times$ and  $\sum_n v^*_{nm'}\int d\omega g^*_{n}(\omega)\times$ on the
left-hand side of Eq. (\ref{eq:cont-eq-cont}), respectively. After this, we can derive the following expressions:
\aln{
0
&= f_{jm}^*(E)-\sum_{j'=1}^D\frac{
A_{j'm}(E)}{E-M_{j'}}\Big(\delta_{j'j}(E-M_{j'})-\sum_{n=1}^C\int_{a_n}d\omega\frac
{f_{j'n}(\omega)f_{jn}^*(\omega)}{E-\omega\pm i0}\Big)
+\sum_{n'=1}^C\bigg(B_{n'm}(E)\sum_{n=1}^Cv_{nn'}\int_{a_n}\frac{f^*_{jn}(\omega)g_n(\omega)}{E-\omega\pm
i0}\bigg)\,,
\label{eq:cont-eq-f}\\
0&= v^*_{mm'}g^*_m(E)+\sum_{j=1}^D\alpha_{jm}\sum_{n=1}^Cv^*_{nm'}
\int_{a_n}\frac
{f_{jn}(\omega)g_{n}^*(\omega)}{E-\omega\pm
i0}-\sum_{n'=1}^CB_{n'm}(E)\Big(v^*_{n'm'}-\sum_{n=1}^Cv^*_{nm'}v_{nn'}\int_{a_n}\frac{g_n(\omega)g^*_n(\omega)}{E-\omega\pm
i0}\Big).
\label{eq:cont-eq-g}}

The  matrix representation proves to be a valuable tool in simplifying the derivation process and achieving concise results. In this context,
we introduce matrix $\mathbf{Y}$ and $\mathbf{F}$ with $(C+D) \times
C$ dimension,
matrices $\mathbf{V}^A$ and $\mathbf{V}^B$, with dimensions $D \times
D$ and $C \times C$ respectively,  matrices $\mathbf{V}^{AB}$ and
$\mathbf{V}^{BA}$, having dimensions $D \times C$ and $C \times D$
respectively,
and  finally, a $(C+D) \times (C+D)$ matrix $\mathbf{M}$
encompassing
$\mathbf{V}^A$, $\mathbf{V}^B$, $\mathbf{V}^{AB}$ and
$\mathbf{V}^{BA}$ as follows
\aln{
(\mathbf
Y)_{jm}=&\alpha_{jm}=\frac{A_{jm}(E)}{E-M_j},\quad  (\mathbf
F)_{jm}= f_{jm}^*,\quad \mathrm{for}\
m=1,\cdots,C;j=1,\cdots,D,
\nonumber\\(\mathbf
Y)_{mn}=&B_{m-D,n},\quad(\mathbf F)_{mn}=v^*_{n,m-D}g^*_n, \quad \mathrm{for}\ n=1,\cdots,C; m=D+1,\cdots,D+C,
\label{eq:Mat-Y-F}
\\
(\mathbf V^A)_{ij}=&\delta_{ij}(E-M_i)-\sum_{n=1}^C\int_{a_n}d\omega\frac
{f_{jn}(\omega)f_{in}^*(\omega)}{E-\omega\pm i0}, \quad \mathrm{for}\
i=1,\cdots,D; j=1,\cdots,D,
\nonumber\\\quad
(\mathbf V^B)_{mn}=&\Big(v_{mn}-\sum_{l=1}^C v_{ml}v_{ln}\int_{a_l}\frac{g_l(\omega)g^*_l(\omega)}{E-\omega\pm
i0}\Big),  \quad \mathrm{for}\
m=1,\cdots,C; n=1,\cdots,C,
,\nonumber\\
(\mathbf V^{AB})_{im}=&-\sum_{n=1}^Cv_{n,m}\int_{a_n}\frac{f^*_{in}(\omega)g_n(\omega)}{E-\omega\pm
i0},  \quad \mathrm{for}\
i=1,\cdots,D; m=1,\cdots,C,
\nonumber\\
 (\mathbf V^{BA})_{mj}=&-\sum_{n=1}^C v^*_{nm}\int_{a_n}\frac{f_{j,n}(\omega)g^*_{n}(\omega)}{E-\omega\pm
i0}, \quad \mathrm{for}\ j=1,\cdots,D; m=1,\cdots,C,
\nonumber\\
\mathbf M_{IJ}=&\begin{pmatrix} \mathbf V^A(E)&\mathbf V^{AB}(E)\\
\mathbf V^{BA}(E)&\mathbf V^{B}(E)\end{pmatrix}_{IJ},  \quad \mathrm{for}\
I,J=1,\cdots,C+D\,.
\label{eq:def-M}}
Similar to the coefficients $\alpha$ and $\psi$, we have omitted
the superscript $\pm$  in the notations of the matrices $\mathbf Y$, $\mathbf V$, and $\mathbf M$, which can be inferred from the surrounding
contexts.
With these matrices, the two equations
(\ref{eq:cont-eq-f}) and (\ref{eq:cont-eq-g}) above can be expressed  in matrix form
\aln {\mathbf M\cdot \mathbf Y=\mathbf F. \label{eq:MY-F}}
or in component form
\aln{
\sum_{j=1}^DV^A_{ij}(E)
\alpha_{jm}(E)+\sum_{n=1}^CV^{AB}_{in}(E)B_{nm}(E)=&f_{im}^*(E)\,,
\label{eq:MY-components-1}\\
\sum_{j=1}^DV^{BA}_{nj}(E)\alpha_{jm}(E)+\sum_{n'=1}^CV^B_{nn'}(E)B_{n'm}(E)=&g^*_{m}(E)\delta_{m,n}\,.
\label{eq:MY-components-2}}

Before further proceeding, let us look at some properties of these
matrices. From the relation $v^*_{mn}=v_{nm}$, we can observe the following symmetric properties:
\aln{
(\mathbf V^{A+})^*_{ij}=&(\mathbf V^{A-})_{ji},\quad (\mathbf V^{B+})^*_{mn}=(\mathbf V^{B-})_{nm},
\quad (\mathbf V^{AB+})^*_{jn}=(\mathbf V^{BA-})_{nj}\,,
\nonumber\\
\mathbf M^{+\dagger}=&\mathbf M^-.
\label{eq:M-hc}
}
In the case where $v_{mn}$, $f_{jm}(\omega)$ and $g_n(\omega)$
are real for real $\omega$, these function matrices possess real
analyticity property. As a result, they can be analytically continued
to the entire complex $E$ plane and satisfy the Schwartz reflection
property.  Moreover, the analytically continued function matrices can
relate the $+i0$ and $-i0$ counterparts, representing the limits
on the upper and lower edges along the real axis above the threshold.
In the cases that $v_{mn}$, $f_{jm}(\omega)$ and
$g_n(\omega)$ are complex, the function matrices will no longer
be real analytic, but the determinant of the matrix
$\mathbf M$, denoted as  $\det
\mathbf M$, remains real analytic.  Thus the analytically continued determinant exhibits the Schwartz reflection symmetry, $\det \mathbf M(z)=\det
\mathbf M^{*}(z^*)$.

In general,  $\det \mathbf M $ is nonzero for general real $E$ values above
the lowest threshold, otherwise if $\det \mathbf M=0 $ for all real
$E$, $v_{mn}$ matrix would be degenerate and there would be a state decouples in
continuum-continuum interaction, which is not our assumption at
present. As a result,  $\mathbf M$ possesses an inverse.   and
$\mathbf Y$ can be obtained by   $\mathbf
Y= \mathbf M^{-1}\cdot \mathbf F$. Then, by inserting $A_{jm}(E)$ or $\alpha_{jm}(E)$ and $B_{nm}$  into Eq.~(\ref{eq:cont-eq-cont}),
 one obtains the coefficients $\psi^\pm_{nm}$,
and the continuum eigenstates are solved to be
\aln{
|\Psi^\pm_m(E)\rangle
=&
|E,m\rangle+\sum_{j=1}^D
\alpha^\pm_{jm}(E)\Big(|j\rangle+\sum_{n=1}^C\int_{a_n} \mathrm d\omega
\frac
{f_{jn}(\omega)}{E-\omega\pm i0}
|\omega;n\rangle\Big)+\sum_{n,n'=1}^C v_{nn'}B^\pm_{n'm}(E)\int_{a_n} \mathrm d\omega
\frac
{g_n(\omega)}{E-\omega\pm i0}
|\omega;n\rangle
}
for $E>a_m$.
Upon comparison with  Eq. (\ref{eq:Cont-m}), this solution is
different  only in the last term, stemming from the presence of separable potential.
Importantly, it can be confirmed that the solution retains the previous normalization condition,  $\langle
\Psi^\pm_m(E)|\Psi^\pm_n(E')\rangle=\delta(E-E')\delta_{mn}$. This normalization condition guarantees the orthogonality of the wave functions, ensuring their compatibility and consistency within the framework of the problem.

The
$S$-matrix can be obtained as
\aln{
S_{mn}(E,E')
=&\delta_{mn}\delta(E-E')-2\pi i\delta(E-E')\Big(\sum_{IJ=1}^{D+C}
(\mathbf F^{\dagger})_{mI}(\mathbf M^+)^{-1}_{IJ}(
\mathbf F_{Jn})\Big)\,,
\label{eq:S-FM-1F}
}
or
\bqa\mathbf S(E,E')
=\mathbf I\delta(E-E')-2\pi i\delta(E-E')
\mathbf F^{\dagger}\cdot (\mathbf M^+)^{-1}\cdot
\mathbf F\,
\eqa in a simplified matrix form.
For a more thorough derivation of the normalization and
meticulous calculation of the $S$-matrix, please refer to the Appendix
\ref{sect:sep-cont:app}, where we provide a detailed presentation of the calculations,
offering a comprehensive and in-depth derivation of the normalization condition and   the $S$-matrix.

Subsequently, our attention turns towards the derivation of  discrete
eigenstates.
The eigenvalues for the discrete states does not coincide with the
spectrum of the continuum states. Thus, using the condition  $E\notin [a_i,\infty)$ for $i=1,\dots, C$,
 Eq.(\ref{eq:eigen-s-eq-2}) one can solve  Eq.(\ref{eq:eigen-s-eq-2})  and obtain
\begin{align}
\alpha_j(E)=&\frac1{E-M_j}A_j(E),
\label{eq:discret-eq-1}
\\
\psi_n(E,\omega)
=&\frac 1{E-\omega}\Big(\sum_{j=1}^D {\alpha_{j}(E)
f_{jn}(\omega)}+\sum_{n'=1}^C v_{nn'}B_{n'}(E)g_n(\omega)\Big)
\,.
\label{eq:cont-eq-1}
\end{align}
By multiplying Eq.~(\ref{eq:cont-eq-1}) with $f^*_{jn}(\omega)$ and
$v^*_{nm}g_n^*(\omega)$ separately, and subsequently
summing over $n$ and integrating
w.r.t. the variable $\omega$, we arrive at the following expressions:
\begin{align}
0
&=-\sum_{j'=1}^D\frac{
A_{j'}(E)}{E-M_{j'}}\Big(\delta_{j'j}(E-M_{j'})-\sum_{n=1}^C\int_{a_n}d\omega\frac
{f_{j'n}(\omega)f_{jn}^*(\omega)}{E-\omega}\Big)
+\sum_{n'=1}^C\bigg(B_{n'}(E)\sum_{n=1}^Cv_{nn'}\int_{a_n}\frac{f^*_{jn}(\omega)g_n(\omega)}{E-\omega}\bigg)\,,
\label{eq:eigen-eq-AB1}\\
0
&=\sum_{j=1}^D\sum_{n=1}^Cv^*_{nm}\frac{ A_{j}(E)}{E-M_j}
\int_{a_n}\frac
{f_{jn}(\omega)g_{n}^*(\omega)}{E-\omega}-\sum_{n'=1}^CB_{n'}(E)\Big(v^*_{n'm}-\sum_{n=1}^Cv^*_{nm}v_{nn'}\int_{a_n}\frac{g_n(\omega)g^*_n(\omega)}{E-\omega}\Big)\,.
\label{eq:eigen-eq-AB2}
\end{align}
The expressions obtained in these two equations deviate from (\ref{eq:cont-eq-f}) and
(\ref{eq:cont-eq-g}) by the absence of the first  terms on the right hand side.
Analogous to the definition in  Eq.~(\ref{eq:def-M}), we can introduce the matrices $\mathbf V^A$, $\mathbf V^{AB}$,
$\mathbf V^{BA}$, $\mathbf V^B$, and $\mathbf M$ as the analytic continuation
of the matrices in Eq.~(\ref{eq:def-M}) and
\[\mathbf X^T=(\frac{A_1}{E-M_1},\dots,\frac{A_D}{E-M_D}, B_1,\dots, B_C)
=(\alpha_1,\dots,\alpha_D,\dots B_1,\dots, B_C)\,.\]
Then
Eqs.(\ref{eq:eigen-eq-AB1}) and (\ref{eq:eigen-eq-AB2}) can be
expressed as
\[\mathbf M\cdot \mathbf X=0.\]
To obtain nonzero solutions for the vector $\mathbf X$, it is essential to satisfy the condition that the determinant of $\mathbf M$  is equal to zero, i.e., $\det \mathbf
M(E)=0$. By analytically continuing this equations to different Riemann
sheets and solving it on each sheet, we can determine  the
generalized discrete eigenvalues.

Once the generalized
eigenvalues are determined, the vector
$\mathbf X$ can be solved for each eigenvalue.
Substituting the solutions for $\alpha_j$ and $B_n$
 into Eq.~(\ref{eq:cont-eq-1}), we
obtain the discrete solution from Eq.~(\ref{eq:expand-2}) for each
generalized energy eigenvalue,
\aln{
|\Psi^{(i)}(E_i)\rangle=\sum_{j=1}^D
\alpha^{(i)}_j(E_i)\Big(|j\rangle+\sum_{n=1}^C\int_{a_n} \mathrm d\omega
\frac{f_{jn}(\omega)}{E_i-\omega}|\omega;n\rangle\Big)+\sum_{n,n'=1}^C
v_{nn'}B^{(i)}_{n'}(E_i)\int_{a_n} \mathrm d\omega
\frac
{g_n(\omega)}{E_i-\omega}
|\omega;n\rangle,
\label{eq:dis-solu}
}
where the superscript $(i)$ denotes the $i$-th discrete solution.
The solution for $\mathbf X$ is only determined up
to a normalization. On the first Riemann sheet, the zeros of
the $\det \mathbf M$ can only be located on the real axis below $a_1$ due to the
hermicity of the Hamiltonian. These zeros correspond to the discrete eigenvalues $E_b$.
It is possible for the associated states to have a finite norm, and we can impose the normalization condition on the coefficients to ensure
\aln{1=&
\sum_{i=1}^D|\alpha_i(E_b)|^2+\sum_{n=1}^C\int_{a_j}^\infty d\omega
\frac 1{(E_b-\omega)^2}\Big|\sum_{j=1}^D {\alpha_{j}(E_b)
f_{jn}(\omega)}+\sum_{n'=1}^C v_{nn'}B_{n'}(E_b)g_n(\omega)\Big|^2
\nonumber\\
=&\mathbf X(E_b)\cdot \mathbf M'(E_b)\cdot \mathbf X^*(E_b)\,.
\label{eq:norm-bound}
}
Within the framework described earlier, each term in the summation can be interpreted as the probability of finding the corresponding bare state within the bound state. However, there could also be complex energy solutions present on different unphysical sheets. As the determinant of $\mathbf M$ is a real analytic function, these complex eigenvalue solutions appear as  complex conjugate pairs.

 As already
mentioned,
there exist $2^C$ distinct Riemann sheets.  However, for our specific purposes, we focus solely on  solutions $E_R$ that reside on the lower half Riemann sheet closest to the physical
sheet. These solutions have a significant impact on the physical $S$-matrix
elements.
Since $E_R$ lies on a nearby unphysical sheet, the evaluation of the matrix value of  $\mathbf M$ at this point requires deforming the integral contours  to the
corresponding sheet around $E_R$ defined in the matrix $\mathbf V$ and
$\mathbf M$  in Eq. (\ref{eq:def-M}) as
illustrated in Fig.~\ref{fig:contour}~\cite{Xiao:2016dsx}.
Also in the state solution Eq.(\ref{eq:dis-solu}), the integral
contours are also deformed similarly. The normalization requirement of these states may resemble Eq.(\ref{eq:norm-bound}), but with $E_b$ replaced by $E_R$, and the
integral contour adjusted accordingly following the deformation depicted in Fig.~\ref{fig:contour}. However, it is important to note that there is no probabilistic explanations for
each terms in the sum, as they may not be real. Additionally, there can
also be real solutions below the lowest threshold $a_1$ on unphysical
sheets, which correspond to virtual states. Similar to the resonant
states, the corresponding integral contours should  be deformed in
Eq.~(\ref{eq:dis-solu}) and in Eq.~(\ref{eq:def-M}) for these states.

 In the case where the coupling constant matrix $v_{mn}$ is
degenerate, certain continuum states may decouple from the contact
interaction. This allow us to choose a suitable set of continuum basis
states in which the decoupled states do not appear in the contact
interaction terms. A more general hamiltonian can be expressed as
\begin{align}
H=&\sum_{i=1}^D M_i|i\rangle\langle
i|+\sum_{n=1}^C \int_{a_n}^\infty \mathrm d \omega
\,\omega|\omega;n\rangle\langle \omega;n|\nonumber
\\ &+\sum_{m,n=1}^r\sum_{m',n'=1}^C v_{mn}\Big(\int_{a_m}^\infty\mathrm d \omega
{ g_{mm'}(\omega)}|\omega;m'\rangle\Big)\Big(\int_{a_n}^\infty\mathrm d \omega
{ g^*_{nn'}(\omega)}\langle \omega;n'|\Big)\nonumber
\\ &+\sum_{j=1}^D\sum_{n=1}^C \left[ |j\rangle\Big(\int_{a_n}^\infty\mathrm d \omega
{ f^*_{jn}(\omega)}\langle \omega;n|\Big)+ \Big(\int_{a_n}^\infty\mathrm d \omega
{f_{jn}(\omega)}|\omega;n\rangle \Big)
\langle j|
\right]\,,
\label{eq:H-degenerate-v}
\end{align}
where $r$ is the rank of the continuum-continuum coupling constant
matrix and $v_{mn}$, $m,n=1,\cdots,r$ is a non-degerate matrix. Notice
that in general, though we are discussing the case for degererate case
where $r<C$, this general interaction even applies for the cases when
$r>C$ which may correspond to the case we will discuss in the next section. Since
$v$ is a hermitian matrix, it can always be diagonalized and be chosen
as $v_{mn}=\lambda_m\delta_{mn}$, for $m,n=1,\cdots,r$. When
$g_{mm'}\propto \delta_{mm'}$, it reduces to the original case~\eqref{eq:H-separable}.
The
solution to the engenvalue problem for this Hamiltonian is
straightforward as before. The difference of the results from the
nondegerate case is roughly to change definition of $B_n$ in
Eq.~\eqref{eq:def-A-B} by $B_n= \sum_{l=1}^C \int g_{nl}\psi_l$
($n=1,\cdots,r$) and
replace the factor $v_{nm} g_n$ to $\sum_{l=1}^r v_{lm}g_{ln}$ in each
equation. For different continuum solution $B_n$ would need another
index $m$ to denote the corresponding continuum solution, i.e. $B_{nm}$,
$(n=1,\cdots,r; m=1,\cdots,C)$. Notice that the range of the first
subindex of $g_{ln}$ and $B_{nm}$ is
from $1$ to $r$ and the second one from $1$ to $C$. Thus the sum of the
first subindex of $g_{ln}$ of $B_{nm}$ need to be  from $1$ to $r$.

A special case is when $r=1$ and we can set $v_{11}=1$ and
$g_{1n}= v_n g_{n}(E)\equiv (\mathbf g)_n$ where
$v_n$ is a constant and $g_n(E)$ is a coupling function. Then the
we can rename $B_{1m}$ to  $B_m^\pm(E)\equiv B^\pm _{1m}=\sum_{n=1}^C v_n \int d\omega g_n(E)\psi^\pm_{nm}(E)$.
The $\mathbf M$ matrix in Eq.~\eqref{eq:def-M} can be represented as
\aln{
\mathbf M_{IJ}=&\begin{pmatrix}\eta& -\mathfrak f
\\-\mathfrak f^{\ddagger T}& \mathfrak g
\end{pmatrix},  \quad \mathrm{for}\
I,J=1,\cdots,D+1\,,
\label{eq:def-vmvn-M}
\\
\eta_{ij}(E)=&\delta_{ij}(E-M_i)-\sum_{n=1}^C\int_{a_n}d\omega\frac
{f_{jn}(\omega)f_{in}^*(\omega)}{E-\omega\pm i0}, \quad \mathrm{for}\
i=1,\cdots,D; j=1,\cdots,D,
\\
\mathfrak g(E)=&\Big(1-\sum_{l=1}^C\int_{a_l}\frac{v_lg_{l}(\omega)v^*_lg^*_{l}(\omega)}{E-\omega\pm
i0}\Big) \,,
\\
\mathfrak f_i=&\sum_{m'=1}^C\int_{a_{m'}}\frac{f^*_{im'}(\omega)v_{m'}g_{m'}(\omega)}{E-\omega\pm
i0},  \quad \mathrm{for}\
i=1,\cdots,D;
\\
\mathfrak f^\ddagger_j=&\sum_{m=1}^C\int_{a_{m}}\frac{f_{j,m}(\omega)v_m^*g^*_{m}(\omega)}{E-\omega\pm
i0},
  \quad \mathrm{for}\
j=1,\cdots,D;
}
For the continuum eigenstate solutions, $\alpha_{jm}$ and $B_m$ are solved  as
\aln{
\alpha_{jm}=&(\eta^{-1}\cdot \mathbf f_m^*)_j+(\eta^{-1}\cdot \mathfrak
f)_jB_m\,,\label{eq:alpha-sol-vmvn}\\
B_m=&\frac{ \mathfrak
f^{\ddagger T}\cdot \eta^{-1}\cdot \mathbf f^*_{m}+(\mathbf g^*(E))_m}{\mathfrak g(E)-\mathfrak f^{\ddagger T} \cdot\eta^{-1}\cdot
\mathfrak f }\,, \quad m=1,\ldots,C,
\label{eq:B-sol-vmvn}
}
where we have defined the vector $(\mathbf f_m)_i=f_{im}$,
$i=1,\cdots,D$.
Then the $S$-matrix can be obtained as
\aln{
S_{mn}(E,E')=&\delta_{mn}\delta(E-E')-2\pi i\delta(E-E')\Big( \mathbf
f_{m}\cdot( \eta^+)^{-1}\cdot \mathbf f^*_n(E)
\\&+\frac{\big(\mathbf
f_{m}\cdot( \eta^+)^{-1}\cdot \mathfrak f^+(E) + (\mathbf g(E))_{m}\big)
 \big(\mathfrak
f^{+\ddagger T}\cdot ( \eta^+)^{-1}\cdot \mathbf f^*_{n}+(\mathbf
g^*(E))_n\big)}{\mathfrak g^+(E)-\mathfrak f^{+\ddagger T} \cdot( \eta^+)^{-1}\cdot
\mathfrak f^+ }\Big)\,.
}
With the $S$ matrix, the observable scattering cross section can be
obtained to compare with the experiments.

In the spirit of effective field theory, a discrete state $j_0$ become
decoupled when its mass $M_{j_0}$ greatly exceeds the system's
characteristic energy scale. This fundamental principle was
exemplified in Weinberg's seminal work~\cite{Weinberg:1962hj}, where
two equivalent formulations were constructed: the full theory contains
the discrete state explicitly in the free Hamiltonian, while the
reduced theory eliminates the discrete state by introducing a specific
potential renormalization. Weinberg demonstrated their equivalence
when $M_{j_0}\to \infty$, provided the potentials satisfy the matching conditions that encode the decoupling dynamics.
 In our present scenario, we can also
construct a
low energy effective Hamiltonian without this discrete state and
include an effective contact interaction of the continuum states after integrate out the
intermediate discrete state in $s$ channel. The corresponding
 interaction term can be expressed using separable contact
effective interaction terms as
\begin{equation}
-\sum_{m,n=1}^C\int d\omega
\frac{f_{j_0m}(\omega)}{\sqrt{M_{j_0}}}|\omega;m\rangle\langle
\omega;n|\frac{f^*_{j_0n}(\omega)}{\sqrt{M_{j_0}}}.
\label{eq:decouple-inter}
\end{equation}
This interaction is similar to the previous special case with $r=1$
by replacing $v_{11}\to v_{j_0j_0}=-1$, and $g_{1n}\to g_{j_0 n}=\frac{f_{j_0
n}}{\sqrt{M_j}}$, and the matrix elements $v_{j_0,n}=0$ and
$g_{mm'}=g_m\delta_{mm'}$.
It effective adds one extra rank to the original $v_{mn}$ matrix.
To solve this eigenvalue problem,
using Eq.~\eqref{eq:def-M} and previous discussion below
\eqref{eq:H-degenerate-v}, there will be a
 row and colomn in the $\mathbf
V^B$ from the \eqref{eq:decouple-inter}, with
\aln{\mathbf
V^B_{j_0j_0}=-1-\sum_{l=1}^C\int_{a_l}\frac{f_{j_0l}(\omega)f^*_{j_0l}(\omega)/M_{j_0}}{(E-\omega\pm
i0)}\,, \mathbf
V^B_{j_0n}=\sum_{l=1}^C\int_{a_l}\frac{v_{ln}g_{l}(\omega)f^*_{j_0l}(\omega)/\sqrt{M_{j_0}}}{(E-\omega\pm
i0)}.\label{eq:VB-j0}}
For $\mathbf V^{AB}$, there will be  corresponding matrix elements
\aln{\mathbf V^{AB}_{ij_0}=&\sum_{n=1}^C\int_{a_n}\frac{f^*_{in}(\omega)f_{j_0n}(\omega)/\sqrt{M_{j_0}}}{E-\omega\pm
i0}\,,\label{eq:VAB-j0}}
and similar for $\mathbf V^{BA}$.
For $\mathbf F$, there is an element $\mathbf F_{j_0,n}=-f^*_{j_0
n}/\sqrt{M_{j_0}}$.
Alternatively, we can start from the original Hamiltonian with the
discrete state and taking the large $M_{j_0}$ limit.
From Eq.
\eqref{eq:def-M}, taking $M_{j_0}$ much larger than $E$ in  $ \mathbf V^{A}_{j_0j_0}$
is to factorize the $M_{j_0}$ and take $E/M_{j_0}\to0$, i.e.  making
the replacement $ \mathbf V^{A}_{j_0j_0}\to
M_{j_0}(-1-\frac 1 {M_{j_0}}
\sum_{n=1}^C\int_{a_n}d\omega\frac{f_{j_0n}(\omega)f^*_{j_0n}(\omega)}{E-\omega\pm
i0})$. Similarly, after factorizing out $M_{j_0}$ from $\mathbf V^A_{j_0
j_0}$,
and $\sqrt{M_{j_0}}$ from
$\mathbf V^A_{ij_0}$, $\mathbf V^A_{j_0i}$, $\mathbf V^{AB}_{j_0m}$, $\mathbf V^{BA}_{mj_0}$, these matrix elements
are of the same form  as the corresponding matrix elements in $\mathbf
V^{AB}$, $\mathbf V^{BA}$, $\mathbf V^B$ in \eqref{eq:VAB-j0} and
\eqref{eq:VB-j0}
as if one is directly solving the decoupled Hamiltonian as constructed above \eqref{eq:decouple-inter}.
One can then find out
that the $S$ matrices obtained by the two approaches are the same.

\section{Approximating a general potential using separable
potentials \label{sect:Approx-Sep}}

In scenarios where the interaction potential can be reasonably
approximated as separable potentials, wherein the potential can be
expressed as the product of two components associated with the ingoing
and outgoing states, respectively, the problem can be effectively
addressed and hold practical significance. Now the problem is how to
approximate a potential using the separable potentials. Before
addressing this problem, let us review how a general contact
interaction can arise in the elastic and nonelastic
scattering.

As discussed in the previous section, the Hamiltonian of a most general model with multiple discrete states and
continuum states and their interactions can be expressed as
\begin{align}
H=&\sum_{i=1}^D M_i|i\rangle\langle
i|+\sum_{n=1}^C \int_{a_n}^\infty \mathrm d \omega
\,\omega|\omega;n\rangle\langle \omega;n|\nonumber
\\ &+\sum_{m,n=1}^C \int_{a_m}^\infty\mathrm d \omega' \int
_{a_n}d \omega
V_{mn}(\omega',\omega)|\omega';m\rangle \langle \omega;n|\nonumber
\\ &+\sum_{j=1}^D\sum_{n=1}^C \int_{a_n}^\infty\mathrm d \omega
\Big(
f^*_{jn}(\omega) |j\rangle\langle \omega;n|+f_{jn }(\omega)|\omega;n\rangle
\langle j|\Big).\nonumber
\end{align}
In the context of nonrelativistic scattering, i.e. when the
in-state and out-state are composed of the same two-particle content,  in
the angular momentum representation, these  continuum states can be expressed as
$|\omega,n\rangle=\sqrt{\mu p}|p,JM;lS\rangle$, where $J,M,l,S$ are the quantum numbers
for total angular
mentum, magnetic angular momentum, relative orbital angular momentum,
and total spin, respectively, and are collectedly denoted using $n$
and $a_n$ are the threshold. Here, $p$ represents the radial
momentum, $\mu$ is the reduced mass and
$\omega=\frac{p^2}{2\mu}+a_n$ represents the total energy. The
normalization of the continuum states is chosen such that the inner
product between two continuum states is given by $\langle
\omega',n'|\omega,n\rangle=\delta_{n'n}\delta(\omega-\omega')$.
The momentum space potential $V_{mn}(\omega',\omega)$ can be derived from the coordinate space potential $V(\blr)$.
For simplicity, we consider only the rotational invariant potential. The potential function $V_{mn}(\omega',\omega)$ arises
from the matrix elements $V_{n'n}(\omega',\omega)\equiv\langle
\omega',n'|V|\omega, n\rangle=\sqrt{\mu' p'}\sqrt{\mu
p}\,\langle p'JMl'S'|V|pJMlS\rangle$.
The simplest example is  when the in-states and out-states are
composed of  the same spinless particles, i.e. elastic scattering. If we know the  coordinate-space
potential $V(r)$, then the momentum-space potential $V(k',k)$ can be expressed in
terms of the coordinate-space potential $V(r)$ as follows:
\begin{align}
\langle k',l',m'|V|k,l,m\rangle=\frac{\delta_{l'l}\delta_{m'm} }{k'k}\int  r^2 \mathrm dr  \frac 2\pi
\hatj_{l}(k'r) V(r)\hatj_{l}(kr),
\end{align}
where the $\hatj(z)$ represents the Riccati-Bessel function.

In the cases where the in-states and out-states can have different particle compositions, we can generalize the potential $V(\omega',\omega)$ accordingly. In addition to the angular momentum quantum numbers, the labels $n$ and
$n'$ can also denote the different particle compositions $|\omega,n\rangle$.
If the potential in coordinate space, $V(\blr',\blr)$, in the
center-of-mass system, is invariant under rotation, it can be expressed as a function of  $\blr^2,\blr'^2$ and $\blr\cdot \blr'$. Here, $\blr$ and $\blr'$ represent
the position of the in-state and out-state relative coordinates,respectively. In the case of spinless particle system,
the matrix elements for in-states and out-states can be expressed as follows:
\als{
V_{n'n}(\omega',\omega)\equiv&\langle n' \omega'lm|V|n,\omega lm\rangle=\int d\blr d\blr'\langle
n',\omega'lm|\blr'\rangle V(\blr',\blr)\langle \blr|n,\omega lm\rangle\nonumber
\\=&\frac2\pi\big(\frac{\mu'\mu}{p'p} \big)^{1/2}\int dr
dr' \hatj_l(pr)\hatj_l(p'r'){\tilde V_{l}(r',r)}\,,
\nonumber\\
\tilde V_{l}(r',r)\delta_{ll'}\delta_{mm'}=&rr'\int d\Omega d\Omega'
Y^*_{l'm'}( \hat\blr')Y_{lm}(
\hat \blr){\tilde V(\blr',\blr)}\,.
}
The Wigner-Ekart theorem has been employed to account for the spherical symmetry of
the potential $V(\blr',\blr)$.
When the in-state and out-state can have spins, the total angular momentum are conserved
but the orbital angular momentum could be different. We can include
the different orbital angular momentum and total spin quantum numbers $lS$
and $l'S'$ into $n$ and $n'$ to label
different in-states and out-states,
\als{
V_{n'n}(\omega',\omega)\equiv&\langle n' \omega'JM|V|n,\omega JM\rangle\nonumber
\\=&\frac2\pi\big(\frac{\mu'\mu}{p'p} \big)^{1/2}\int dr
dr'\sum_{ll'SS'} \hatj_l(pr)\hatj_{l'}(p'r'){\tilde
V_{ll'SS'}^{JM}(r',r)},
\\
\tilde V_{ll'SS'}^{JM}(r',r)=&rr'\sum_{mm'm_sm_s'}\int d\Omega d\Omega'
Y^*_{l'm'}( \hat\blr')Y_{lm}(\hat
\blr){\tilde V_{SS'}(\blr',\blr)}C^{JM}_{lmSm_s}C^{JM}_{l'm'S'm_s'}\,,
}
where $C^{JM}_{lmSm_s}$ is the Clebsch-Gordon coefficients.
To make further progress, we also suppose that the potential is square
integrable for both $\omega'$ and $\omega$, and the same for the
interaction vertex between the discrete states and the continuum
states $f_{jn}(\omega)$, that is
\[\int_{a_m}d\omega'\int_{a_n}d\omega
|V_{mn}(\omega',\omega)|^2=\mathrm{finite}\,,\quad \int_{a_n}d\omega
|f_{jn}(\omega)|^2=\mathrm{finite}.\]

There are no exact solutions for Hamiltonian for general potentals
$V_{mn}(\omega',\omega)$. However, it is well-known that  such a
potential can be expanded using a sum of separable
potentials~\cite{Newton:1982qc,Sitenko:2003}.
For continuum states $|\omega,m\rangle$, we can choose a set of
complete basis functions
$\tilde g_{m\rho}(\omega)$, with $\int_{a_m} d\omega \tilde
g^*_{m\rho}(\omega)\tilde
g_{m\delta}(\omega)=\delta_{\rho\delta}$ and $\sum_\delta \tilde
g_{m\delta}(\omega')\tilde
g_{m\delta}(\omega)=\delta(\omega'-\omega)$.  The basis sets for different continuum
states, i.e. for different $m$ and $n$, do not need to be the same. The potential
$V_{mn}(\omega',\omega)$
can then be expanded as
\[V_{mn}(\omega',\omega)=\sum_{\rho\delta} v_{mn,\rho\delta }\tilde
g_{m\rho}(\omega)\tilde g^*_{n\delta}(\omega).\]
In the following we will use the Greek letter $\rho$, $\delta$ to
label the basis, and repeated Greek letters are summed over without
explicit sum symbol and the sum
symbol for the Latin letters would still be left explicit.
The coefficient matrix composed of $v_{mn,\rho\delta}$ is hermitian
$v_{mn,\rho\delta}=v^*_{nm,\delta\rho}$ and is supposed to be
non-degenerate.
In general, there are infinite number of bases, the sum of $\delta$ and
$\rho$ is up to infinity. Since the expansion coefficients
$v_{mn,\rho\delta}$ are small at large enough order, one can make an approximation  and
truncate the series to a finite order $N$, i.e. $v_{mn,\rho\delta}=0$
for $\rho,\delta>N$.
Then, the general Hamiltonian for multiple continuum states and discrete
states can be recast as
\begin{align}
H
=&\sum_{i=1}^D M_i|i\rangle\langle
i|+\sum_{n=1}^C \int_{a_n}^\infty \mathrm d \omega
\,\omega|\omega;n\rangle\langle \omega;n|
\nonumber\\ &+\sum_{m,n=1}^C v_{mn,\rho\delta}\Big(\int_{a_m}^\infty\mathrm d
\omega'
 \tilde g_{m\rho}(\omega')|\omega';m\rangle\Big)\Big(\int_{a_n}^\infty\mathrm d \omega
 \tilde g^*_{n\delta}(\omega)\langle \omega;n|\Big)
\nonumber\\ &+\sum_{j=1}^D\sum_{n=1}^C \left[ |j\rangle\Big(\int_{a_n}^\infty\mathrm d \omega
 f^*_{jn}(\omega)\langle \omega;n|\Big)+\Big(\int_{a_n}^\infty\mathrm d \omega
 f_{jn} (\omega)|\omega;n\rangle \Big)
\langle j|
\right]\,.
\label{eq:H-Approx-Sep}
\end{align}
One can take $m\rho$ and $n\delta$ as the row and column indices and
diagonalize the matrix $v_{mn,\rho\delta}$. Then, the problem reduces to
the similar case in
 the last part of the last section when the degenerate $v_{mn}$ is
discussed. Here the rank of $v_{mn,\rho\delta}$ is greater than
$C$ just as we mentioned in last section.
Alternatively, we can also directly solve the problem as before in the
following.
The general eigenstate for this eigenvalue problem can be expanded using
the bare discrete states and the bare continuum states
\begin{align}
|\Psi(E)\rangle
=&\sum_{i=1}^D
\alpha_i(E)|i\rangle+\sum_{n=1}^C\int_{a_n} \mathrm d\omega
\psi_{n}(E,\omega)|\omega;n\rangle.
\label{eq:expand-3}
\end{align}
Similar to previous sections, for $|\Psi_m\rangle$, the corresponding $\alpha_i$ and $\psi_n$ will have
another index $m$, i.e. $\alpha_{im}$ and $\psi_{nm}$.

With the same procedures as the previous section, the approximate
properly normalized continuum states
can be solved as
\begin{align}
|\Psi^\pm_m(E)\rangle
=&|E;m\rangle+\sum_{j=1}^D
\alpha^\pm_{jm}(E)\Big(|j\rangle+
\sum_{n=1}^C\int_{a_n} \mathrm d\omega\frac{f_{jn}(\omega)}{E-\omega\pm i0}
|\omega;n\rangle\Big)\nonumber\\
+&\Big( \sum_{n,n'=1}^Cv_{nn',\delta'\rho}\psi^\pm_{n'm\rho}(E)\int_{a_n} \mathrm d\omega\frac{\tilde
g_{n\delta'}(\omega)}{E-\omega\pm i0}
|\omega;n\rangle\Big),
\end{align}
where $\alpha^\pm_{im}(E)$ and
$\psi^\pm_{n'm\rho}(E)\equiv\int_{a_n}^\infty \psi_{n'm}(\omega)\tilde
g_{n\rho}(\omega)$ can be solved as
in Eq.(\ref{eq:tildY-solu}).
The $S$-matrix can then be obtained,
\aln{
\langle
\Psi_m^-(E)|\Psi_n^+(E')\rangle=&\delta_{mn}\delta(E-E')
-2\pi i\delta(E-E')\Big(\tilde{\bf F}^\dagger_m \cdot (\tilde{\bf
M}^+)^{-1}\cdot \tilde{\bf F}_m
\Big)
}
where the matrix $\tilde{\mathbf M}^+$ with dimension
$(D+NC)\times(D+NC)$ and vector $\tilde {\mathbf
F}_m$ with dimension $(D+NC)$ are defined in Eq.(\ref{eq:tilde-Y} ).
The detailed calculation is left to appendix~\ref{sect:solu-Approx-4p}.
The  discrete eigenvalues can be obtained by solving Eq. $\det \tilde
{\mathbf M}(E)=0$. The discrete
state corresponding to eigenvalue $E_i$ can be expressed as
\begin{align}
|\Psi^{(i)}(E_i)\rangle
=&\sum_{i=1}^D
\alpha^{(i)}_{j}(E_i)\Big(|j\rangle+
\int_{a_n} \mathrm d\omega\frac{f_{jn}(\omega)}{E_i-\omega}
|\omega;n\rangle\Big)
+\sum_{n'=1}^C
v_{nn',\delta'\rho}\psi^{(i)}_{n'm\rho}(E_i)\int_{a_n} \mathrm d\omega\frac{\tilde
g_{n\delta'}(\omega)
}{E_i-\omega}
|\omega;n\rangle\,,
\end{align}
where  the integral contour needs to be deformed for $E_i$ on
unphysical sheets as before.

It is worthful to mention that though we are choosing $g_{m\delta}$ as orthogonal
function sets, we do not use this orghogonal property in solving the
problem. Thus, as long as we can approximate the Hamiltonian using the
separable interaction like in \eqref{eq:H-Approx-Sep} without
orthogonal conditions for $g_{m\delta}$ functions, the solution applies.


Next, we could go further and
also expand the interaction function $f_{jm}(\omega)$ using the same
set of basis $\tilde g_{m\delta}(\omega)$ as in the corresponding contact interaction involving the
 continuum state $|\omega,m\rangle$,
\[f_{jm}(\omega)=\sum_{\delta}f_{jm\delta}\tilde
g_{m\delta}(\omega),\quad f_{jm\delta}=\int d\omega\, f_{jm}(\omega)\tilde
g^*_{m\delta}(\omega)\,,\]
 and also make an
approximation by truncating the series to the $N$-th order the same
as in the contact terms, that is,
$f_{jm\delta}=0$ for $\delta>N$. This may reduce the dimension of the
matrix $\tilde{\mathbf M}$ and may also simplify the numerical calculation.
Then, the general Hamiltonian for
multiple continuum states and discrete states can be recast as
\begin{align}
H=&\sum_{i=1}^D M_i|i\rangle\langle
i|+\sum_{n=1}^C \int_{a_n}^\infty \mathrm d \omega
\,\omega|\omega;n\rangle\langle \omega;n|
\nonumber \\ &+\sum_{m,n=1}^C v_{mn,\rho\delta}\Big(\int_{a_m}^\infty\mathrm d
\omega'
\tilde g_{m\rho}(\omega')|\omega';m\rangle\Big)\Big(\int_{a_n}^\infty\mathrm d \omega
 \tilde g^*_{n\delta}(\omega)\langle \omega;n|\Big)
\nonumber \\ &+\sum_{j=1}^D\sum_{n=1}^C \left[f^*_{jn\delta} |j\rangle\Big(\int_{a_n}^\infty\mathrm d \omega
\tilde g^*_{n\delta}(\omega)\langle \omega;n|\Big)+ f_{jn\delta } \Big(\int_{a_n}^\infty\mathrm d \omega
\tilde g_{n\delta}(\omega)|\omega;n\rangle \Big)
\langle j|
\right]\,.
\label{eq:H-Approx-Sep-f}
\end{align}
This case is more like the cases discussed
in~\cite{Xiao:2016mon,Sekihara:2014kya}, where the same form factor
comes with the continuum both in the discrete-continuum and
continuum-continuum interaction.
Using the eigenstate ansatz
\begin{align}
|\Psi(E)\rangle
=&\sum_{i=1}^D
\alpha_{i}(E)|i\rangle+\int_{a_n} \mathrm d\omega
\psi_{n}(E,\omega)|\omega;n\rangle
\nonumber \\
=&\sum_{i=1}^D
\alpha_i(E)|i\rangle+\sum_{n=1}^C\psi_{n\delta}(E)\int_{a_n} \mathrm d\omega
\tilde g_{n\delta}(\omega)|\omega;n\rangle\,,
\label{eq:expand-4}
\end{align}
one can  proceed solving the eigenvalue problem similarily to the
previous section, of which the details will be left to appendix~\ref{sect:solu-Approx-3p}.
The properly normalized continuum state can be solved and  expressed as
\begin{align}
|\Psi^\pm_m(E)\rangle
=&\sum_{i=1}^D
\alpha^\pm_{im}(E)|i\rangle+|E,m\rangle+\sum_{n=1}^C\sum_{n'=1}^C\psi^\pm_{n'm\rho}(E)
V_{n\delta',n'\rho}(E)\int_{a_n} \mathrm d\omega
\frac{\tilde
g_{n\delta'}(\omega)}{E-\omega\pm i0}
|\omega;n\rangle
\label{eq:expand-Ext-sol}
\end{align}
where the $\alpha^\pm_{im}$ and $\psi^\pm_{nm\rho}$ can be solved from
Eq.(\ref{eq:psi-tilde}).
The $S$-matrix can be obtained
\aln{
\langle
\Psi_m^-(E)|\Psi_n^+(E')\rangle=&\delta_{mn}\delta(E-E')-2\pi
i\delta(E-E')\sum_{n=1}^C\sum_{n'=1}^C\psi^{-*}_{n'm\rho}(E')
V^*_{n\delta',n'\rho}(E')
\tilde
g^*_{n\delta'}(E)
\nonumber \\=&\delta_{mn}\delta(E-E')-2\pi
i\delta(E-E')(\tilde {\mathbf F}_m^\dagger\cdot (\tilde {\mathbf W}^{+})^{-1}
\cdot
\tilde {\mathbf F}_n)
}
where the $NC\times NC$ matrix $\tilde {\mathbf W}^+$ and $NC$
dimensional vector $\tilde {\mathbf F}_m$
are defined in Eq.~(\ref{eq:tilde-W}) and Eq.~(\ref{eq:tilde-F-Y}).
Similar to previous section,
the generalized energy eigenvalues for the discrete state can be
obtained from the $\det \tilde {\mathbf M}(E)=0$, and for each
eigenvalue $E_i$, $\psi_{n\rho}^{(i)}$ can be solved from $\tilde
{\mathbf M}\cdot \tilde {\mathbf Y}=0$, where the matrix $\tilde
{\mathbf M}$ and $\tilde {\mathbf Y}$ are defined in
Eq.~(\ref{eq:M-Ext})
and Eq.~(\ref{eq:tilde-F-Y}). Then
we have
the discrete eigenstates,
\aln{
|\Psi^{(i)}(E_i)\rangle=&\sum_{n'=1}^C\psi^{(i)}_{n'\rho}(E_i)\Big[\sum_{j=1}^D
\frac{f^*_{jn'\rho} }{E_i-M_j}
|j\rangle+V_{n\delta',n'\rho}(E_i)\int_{a_n}d\omega\frac{\tilde
g_{n\delta'}(\omega)}{E_i-\omega}
|\omega;n\rangle\Big]
\\
\mathrm{with \,}& \tilde {\mathbf Y}^{(i)\dagger}(E_i)\cdot  \mathbf V(E_i)\cdot
\tilde{\mathbf W}'(E_i)\cdot \mathbf V(E_i)\cdot  \tilde {\mathbf Y}^{(i)}(E_i)=1
\nonumber
}
with integral contour deformed for resonances and virtual states as before.
\section {Application in analyzing the discrete state position under
interactions\label{sect:discuss}}
In the case when there are contact interactions involved, we can
explore the effect of introducing small couplings
on the
mass of the discrete states in a general manner. Analyzing pole trajectories as couplings vary offers valuable insights  into particle properties. This type of analysis proves useful in elucidating the origin of certain states observed in the experiments utilizing Friedrichs-like models or similar formulas derived from the dispersive models.
 Notably, refs.~\cite{Zhou:2010ra, vanBeveren:2006ua,
Giacosa:2019ldb,Albaladejo:2016lbb,Meissner:2020khl} demonstrate how pole trajectories
of various states provide valuable clues regarding the possible nature of these particles.
Moreover, such analyses may also provide qualitative
guidance in understanding the interaction properties from the
spectrum.  To illustrate this, we focus on the exponential form factor, a frequently employed form factor in the literature. Using this form factor as an example, we  discuss the properties of the bound
states while varying the couplings. For simplicity, we consider a
two-channel case where the threshold for the two continuum states are
denoted as $a_1$ and $a_2$ with $a_1<a_2$.

\begin{figure}
\includegraphics[height=2cm]{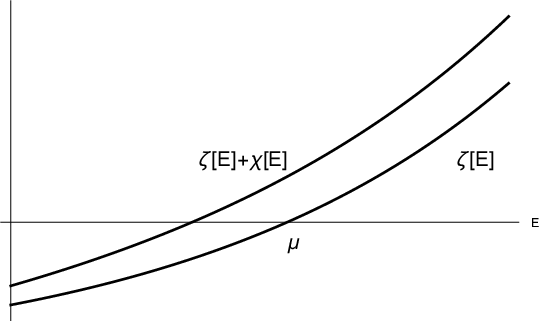}
\caption{A monotonic increasing function $\zeta(E)$ which satisfies
$\zeta(\mu)=0$. The solution for $\zeta(E)+\chi(E)=0$ is smaller than $\mu$
if $\chi(E)$ is a monotonic increasing positive
function.\label{fig:illus}}
\end{figure}

 The basic consideration is as follows: At
the leading order, where the interactions are absent, the discrete state is
determined by the condition $E-\mu=0$, with the bare mass being the solution. When  a small coupling constant $\lambda$ is turned on, we must consider an
equation of the form $E-\mu+\lambda \chi(E)=0$ ($\lambda>0$), where
$\chi(E)$ is real and small in the vicinity of  $E=\mu$. The next-to-leading-order
solution can be expressed as $E=\mu-\lambda \chi(\mu) +O(\lambda^2)$.
The sign of $\chi(\mu)$ allow us to determine the tendency of the
solution's behavior.  If we know that $\chi(E)$ is a monotonic
function, either positive definite or
negative definite, we can  determine the direction in which the solution will shift as $\lambda$ increases continuously. For example, if $\chi(E)$ is a positive decreasing function, it is evident that the solution will
move downward as $\lambda$ becomes increasingly positive. More generally, when we replace the left-hand
side of the leading-order equation $E-\mu=0$ with another increasing
function, i.e. $\zeta(E)=0$, the addition of another positive increasing function will
cause the zero point to shift to the left.  See
Fig.~\ref{fig:illus} for an illustration.

\begin{figure}
\includegraphics[height=3cm]{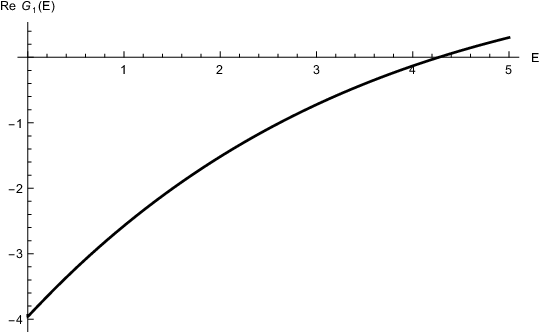}
\quad\includegraphics[height=3cm]{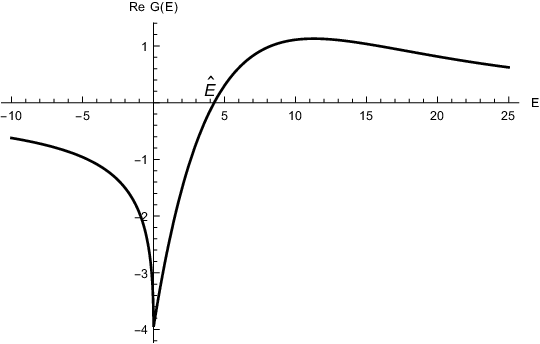}
\caption{The general behavior of the real part of $ G(E)=\int_{a}^\infty d\omega
\frac{|f(\omega)|^2}{E-\omega+i\epsilon}$ function,
 using  $|f(\omega)|^2=\sqrt{\omega-a}
e^{-\omega/\Lambda}$, with $\Lambda=5$, $a=0$. If the interested
energy region is much smaller than $\Lambda$, then $\mathrm {Re} G(E)<0$.
\label{fig:G1}
}
\end{figure}

We also need to examine the behavior of the dispersion integral defined as
$G(E)=\int_a^\infty d\omega \frac{ |f(\omega)|^2}{E-\omega +i0}$. When
$E<a$, the integral behaves as a purely negative decreasing function. For $E>a$, the
imaginary part is $-\pi |f(E)|^2$, which is purely negative in this region.
The real part corresponds to the principal value integral in $G(E)$.  Near the threshold, the real
part is negative and increasing with $E$, passing
through zero point $\hat E$ and reaching a positive maximum, and then decreases back
towards zero as $E$ approaches infinity. See Fig.  \ref{fig:G1}
for an example.  Typically, the integrand includes a phase space factor
$\rho(E)\propto \sqrt{E-a_1}$ which suppresses the integrand near the
threshold $a$, which causes a higher value of $\hat E$.  See Fig. \ref{fig:G1}
for the case with $|f(\omega)|^2=\sqrt{\omega-a_1} e^{-E/\Lambda}$. We
observe that $\mathrm{Re}G(E)$
becomes positive only when $E$ approaches $\Lambda$, which
characterizes the inverse of the interaction range. If the energy range of interest is significantly smaller than
$\Lambda$, then $\mathrm{Re} G(E)$ remains negative. We will primarily focus on this region below.

With this preparation, we can explore some interesting and instructive simple cases that are relevant to the phenomenological analysis of the spectrum.

First, we consider the cases with only continuum states.
\begin{enumerate}
\item In the presence of a single continuum and self interaction $v_{11}$, if $v_{11}$
is sufficiently negative (i.e. attractive), a bound state will emerge  at $E_0<a_1$.

This is the simplest case. The $\mathbf M$ matrix reduces to a
function,
\[M_{11}=v_{11}(1-v_{11}G_1(E)),\]
where $G_1=\int_{a_1}^\infty d\omega
\frac{|g_1(\omega)|^2}{E-\omega+i\epsilon}$ and $a_1$ is the threshold.
Since $G_1(E)$ is negative and continuously decreasing for $E<a_1$ with $\lim_{E\to -\infty} G_1(E)=0$, as shown in Fig.
\eqref{fig:G1}, for $M_{11}$ to have a zero
point  at $E_0<a_1$, the coupling
$v_{11}$ must satisfy the condition $v_{11}< 1/G_1(a_1)$.

\item \label{it:2-cont-1-bound}
 With a second continuum state included, we can examine the coupled-channel effect on the dynamical bound state of the
lower channel discussed in previous case when $v_{11}<0$.

The $\mathbf M$ matrix becomes
\begin{align}
\mathbf
M=&\begin{pmatrix}v_{11}(1-v_{11}G_1(E))-|v_{12}|^2G_2(E)&v_{12}(1-v_{11}G_1(E))-v_{12}v_{22}G_2(E)
\\
 v_{21}(1-v_{11}G_1(E))-v_{22}v_{21}G_2(E)&v_{22}(1-v_{22}G_2(E))-|v_{12}|^2G_1(E)
\end{pmatrix},
\\ G_i=&\int_{a_i}^\infty d\omega
\frac{|g_i(\omega)|^2}{E-\omega+i\epsilon},\nonumber
\end{align}
where $a_1$ and $a_2$ are the thresholds for the two channels with
$a_1<a_2$. The determinant can be calculated to be
\[\det \mathbf M=
(\det \mathbf v)\Big(1-v_{11}G_1(E)-v_{22}G_2(E)+(\det \mathbf v)
G_1(E)G_2(E)\Big)\,,\quad
\mathbf v=\begin{pmatrix} v_{11}&v_{12}\\v_{21}&v_{22}\end{pmatrix}\,.\]
It is worth mentioning that since $v_{12}$
appears only in $\det \mathbf v$ in the form of $|v_{12}|^2$, the phase or sign of
$v_{12}$ is irrelavent to this result in this context.

There are four cases to consider progressively. In each case we will first present
the result and then provide the reasoning behind it.
\begin{enumerate}
\item When $v_{12}\neq 0$ and $v_{22}=0$, the coupled channel effect will play
a role of attractive interaction, causing the bound state to shift from $E_0$
to a  deeper energy level $E_{0b}$, i.e. $E_{0b}<E_0$.

This can be demonstrated by directly calculating the determinant of  $\mathbf M$, yielding
\aln{\det \mathbf
M=-|v_{12}|^2\Big(1-v_{11}G_1(E)-|v_{12}|^2G_1(E)G_2(E)\Big)\,.
\label{eq:detM-v220}}
At $E<a_1$, the last term
$-|v_{12}|^2G_1(E)G_2(E)$ in the bracket is negative and decreasing, while the term $-v_{11}G_1(E)$ is also negative and decreasing.
Following similar reasoning as illustrated in Fig.~\ref{fig:illus},
$E$ must be much smaller in order to have a
zero point of $\det\mathbf M$, denoted as $E_{0b}$. This discussion
does not rely on the smallness of the coupling, and therefore it is also valid for
strong interaction.

\item When a sufficiently small $v_{22}$ is then turned on, the result
depends on the sign of $v_{22}$. A negative $v_{22}$ causes the bound
state to shift to a deeper energy level, while a positive $v_{22}$ results in a shallower bound state.

The contribution from $v_{22}$ to the $\det \bf M$
at the zero point $E_{0b}$ can be expressed as
\[-v_{22}|v_{12}|^2(v_{11}v_{22}-|v_{12}|^2)G_1(E_{0b})G^2_2(E_{0b})\sim
v_{22}|v_{12}|^4G_1(E_{0b})G^2_2(E_{0b})+O(v_{22}^2)\,.
\]
 When the $v_{22}$ is
negative, indicating a more attractive interaction, this term contributes positively
to $\det \mathbf M$ for $E<a_1$, thereby negatively affecting the terms in the
bracket of~\eqref{eq:detM-v220}. This results in the bound state
shifting deeper from $E_{0b}$,  moving away from the threshold. Conversely, when $v_{22}$ is positive, the bound state becomes shallower, moving upward  from $E_{0b}$. Therefore, when both  $v_{12}$ and a positive $v_{22}$ are present,
the direction of  the bound state's movement  from $E_0$ depends  on
the competition between the effects of $v_{22}$ and $v_{12}$.

\item For large $|v_{22}|$, when $v_{22}>0$, we can conclude that the
bound state will be located to the left of $E_0$ but to the right of
$E_{0b}$.

This can be understood by
expressing $\det \mathbf M $ as follows:
\aln{\det \mathbf M=
(\det \mathbf
v)(1-v_{22}G_2(E))\Big(1-v_{11}G_1(E)-\frac{|v_{12}|^2}{1-v_{22}G_2(E)}
G_1(E)G_2(E)\Big)
\label{eq:detM-v22-1}}
and by considering the positivity of $-v_{22}G_2(E)$ in the last term in the bracket for
$E<a_1$. As $v_{22}$ increases, the absolute value of the last term
in the bracket deceases. Thus, when positive $v_{22}$ is turned on from $0$
to $\infty$, the bound state moves from $E_b$ to $E_0$.

\item
When negative $v_{22}$ is introduced, starting from zero and becomes increasingly negative, the bound
state will shift deeper from $E_{0b}$ to $-\infty$.

This occurs because, for fixed $E$, the term $1-v_{22}G_2(E)$ in the denominator of
the last term in the bracket approaches zero, leading to a divergence of that term. To cancel the first two finite terms, $E$ must become increasingly negative so that $|G_2(E)|$ decreases sufficiently, allowing the last term to remain finite and effectively cancel the first two
terms in the bracket. Consequently, this will cause the bound state to shift deeper
from $E_{0b}$ to $-\infty$.

\end{enumerate}

\item
Similar to previous case, when there is only an
attractive interaction in  the
second channel, i.e. $v_{22}<0$ with $v_{11}=v_{12}=0$,
a dynamically generated bound state ($E_b$) can exist below
$a_2$ and is assumed to be above $a_1$, satisfying the condition
$1/G_{22}(a_1)<v_{22}<1/G_{22}(a_2)$.
We can then examine the effect of turning on the first channel on the
bound state spectrum of
the second channel. There are two cases to consider.
\begin{enumerate}

\item   When
a small $v_{12}$ is introduced while keeping  $v_{11}=0$,
the bound state will transit to the second sheet. Whether the mass of the
state increases or decreases depends on the sign of $\mathrm{Re} G_1(E_b)$. A
negative  $\mathrm{Re} G_1(E_b)$ will cause the mass of the
state tend to decrease, while a positive $\mathrm{Re} G_1(E_b)$ will lead to an increase in the mass.

In this case, we have
\aln{\det \mathbf
M=-|v_{12}|^2\Big(1-v_{22}G_2(E)-|v_{12}|^2G_1(E)G_2(E)\Big)\,.
\label{eq:detM-v22}
}
Consider solving $\det M=0$ in the powers of $v_{12}$ using iteration.
Since $E_b>a_1$, the $G_1(E)$ factor in the last term contributes an
imaginary term of $-\pi i |g_1(E)|^2$. Combining this with the other factors
yields a negative imaginary part on the order of  $|v_{12}|^2$ in the
bracket. For $\det M=0$ to hold, the state corresponding to $E_b$ must move into the complex plane, allowing the term $-v_{22}G_2(E)$ to generate a negative
imaginary part of order $v^2_{12}$ to cancel the
imaginary part of the last term in the bracket. Thus, $E$ must
have a negative imaginary part, and with the
$+i\epsilon$ in the
definition of $G_i(E)$, the pole moves continuously to the second
Riemann sheet.  This is  consistent with the common knowledge that
the resonance poles can not reside on the physical sheet.
Whether the mass of the pole is increasing or decreasing depends on the
 the sign of the leading real part of the last term, which is
determined by examining
$\mathrm{Re} G_1(E_b)=P.V.\int_{a_1}^\infty d\omega
\frac{|g_1(\omega)|^2}{E_b-\omega}$. If this value is
negative(positive) the mass will decrease(increase). This is due to $G_2(E)$ being
negative and decreasing for $E<a_2$, leading to a negative (positive) contribution from the real part of the last term
in the bracket.
In the example illustrated in
Fig.~\ref{fig:G1}, when
$\Lambda$ is much larger than $a_2$
where $\mathrm{Re} G_1(E)$ is negative and  monotonically increasing near
$E_b$,
 turning on the $v_{12}$ will cause the mass of the dynamically
generated state (from $E_b$) in the second channel to decrease.

\item When a small $|v_{11}|$ is also
turned on, as long as $|v_{11}|$ is small enough,  the result will be
the same as previous case.

Similar to previous case,  $\det \mathbf M$ can be
expressed as
\aln{\det \mathbf M=&
(\det \mathbf
v)(1-v_{11}G_1(E))\Big((1-v_{22}G_2(E)-\frac{|v_{12}|^2}{1-v_{11}G_1(E)}
G_1(E)G_2(E)\Big)
\nonumber\\
=&
(\det \mathbf
v)(1-v_{11}G_1(E))\Big((1-v_{22}G_2(E)-\frac{|v_{12}|^2\big(1-v_{11}(\mathrm{Re}
G_1(E)-i \mathrm{Im} G_1(E))\big)}{|1-v_{11}G_1(E)|^2}
G_1(E)G_2(E)\Big)
}
Compared to \eqref{eq:detM-v22}, there is an extra factor  $1/(1-
v_{11}G_1)$ in the
$|v_{12}|^2$ term. The difference is of order $O(v_{11}|v_{12}|^2)$, and
therefore,
for sufficiently small  $|v_{11}|$, the result would be the same as in the previous
case.
\end{enumerate}

\item \label{it:v12-bound}  There could
also be a bound state generated from pure $v_{12}$ interaction with no
self-interaction, i.e. $v_{11}=v_{22}=0$, regardless of the sign of
$v_{12}$. In this case, $v_{12}$ acts as an attractive interaction.

 If $v_{11}=v_{22}=0$, $\det \mathbf
M=-|v_{12}|^2(1-|v_{12}|^2G_1(E)G_2(E))$. Since both $G_1(E)$ and
$G_2(E)$  are negative and decreasing for $E$ below the first threshold, there can
be a solution to $\det \mathbf M=0$ when $|v_{12}|^2$ is sufficiently large,
specifically when $|v_{12}|^2>1/(G_1(a_1)G_2(a_1)))$.
When $|v_{12}|^2$ decreases, the bound state will go up through
the threshold to the second Riemann sheet, becoming a virtual state
or a resonance.
Thus, activating only $v_{12}$ is equivalent to enhance the attractive
interaction when there are only two continuum states.

\end{enumerate}

Next, we add a discrete state of bare mass $\mu$ with coupling vertex
functions to the two continua, $f_1(\omega)$ and $f_2(\omega)$.
Now $\mathbf M$ matrix becomes
\begin{align}
\mathbf M=&\begin{pmatrix}
(E-\mu)-\mathcal F_1(E)-\mathcal F_2(E)& -v_{11}\mathcal F^g_{1}(E)-v_{21}\mathcal
F^g_{2}(E)&- v_{12}\mathcal F^g_{1}(E)-v_{22}\mathcal F^g_{2}(E)
\\
-v_{11}\mathcal F^{g\ddag}_{1}(E)-v^*_{21}\mathcal
F^{g\ddag}_{2}(E)&v_{11}(1-v_{11}G_1(E))-|v_{12}|^2G_2(E)&v_{12}(1-v_{11}G_1(E))-v_{12}v_{22}G_2(E)
\\
-v^*_{12}\mathcal F^{g\ddag}_{1}(E)-v_{22}\mathcal
F^{g\ddag}_{2}(E)&v_{21}(1-v_{11}G_1(E))-v_{22}v_{21}G_2(E)&v_{22}(1-v_{22}G_2(E))-|v_{12}|^2G_1(E)
\end{pmatrix}
\\
\mathrm{where}\quad\mathcal F_{n}=&\int_{a_n}d\omega\frac
{f_{n}(\omega)f_{n}^*(\omega)}{E-\omega+ i0},\quad \mathcal
F^g_{n}=\int_{a_n}d\omega\frac
{f^*_{n}(\omega)g_{n}(\omega)}{E-\omega+ i0},\quad
\mathcal F^{g\ddag}_{n}=\int_{a_n}d\omega\frac
{f_{n}(\omega)g^*_{n}(\omega)}{E-\omega+ i0}.
\nonumber
\end{align}
We look at three different cases.
\begin{enumerate}
\item One bare discrete state coupled with two continuum states with
$v_{ij}=0$.

When there is only the interaction between the discrete state and
the continuum states, i.e.
$v_{11}=0, v_{12}=0, v_{22}=0$, we need only to look at
\aln{M_{11}\equiv(E-\mu)-\mathcal F_1(E)-\mathcal F_2(E)=0.}
By default, we will suppose $\mu\ll \Lambda$, and $\mathrm {Re}
\mathcal F_1(\mu)<0$ similar to Fig. \ref{fig:G1}.
\begin{enumerate}
\item
When $\mu<a_1$, since $\mathcal F_1(E)$ and $\mathcal F_2(E)$ for
$E<a_1$ are negative and decreasing, the solution, denoted as $E_\mu$,
will be less than $\mu$. This indicates that when the discrete state
lies below both continuum states, turning on the interaction between the
discrete state and the continuum states causes the discrete state move
deeper below the thresholds.
\item
For $a_1<\mu<a_2$, when the sufficiently weak interactions $f_1$ and
$f_2$ are gradually
turned on, the discrete state would move to the second sheet and the
mass will go down.

In this case, only $\mathcal F_{1,2}$ contributions near $\mu$ are
significant to the shift of the zero point, as observed in the iteration
solution.
With $\mathcal F_2(\mu)<0$, and assuming $\mathrm{Re} \mathcal F_1(\mu) <0$ (as seen in the case of exponential form factor with
$\mu\ll \Lambda$), a weak interaction between the discrete state and the
two continua will also lead to a decrease in the mass of the discrete state. Since
$\mathcal F_1(\mu)$ has a negative imaginary part, the energy of the discrete state
solution and $-\mathcal F_2(E)$ will develop  negative imaginary parts to cancel it. As a result,
the discrete state
will move continuously to the complex plane of the second Riemann sheet.
\item When $\mu>a_2$, similar to the previous case, the discrete state
will go to the third Riemann sheet and the mass will decrease.

The discussion is similar  to the previous item. When $\mu\ll
\Lambda$, we find that $\mathrm {Re}(\mathcal
F_{1,2}(\mu)) <0$, and these terms act as attractive interactions, driving
the mass of the discrete state downward. The negative imaginary parts
of  $\mathcal F_{1,2}(\mu)$ result in the solution moving down to the third
sheet of the Riemann surface.
\end{enumerate}
This may provide an understanding of why most of the open-flavor effects
tend to cause the mass of the $c\bar c$ state
smaller~\cite{Zhou:2011sp}: The
$\Lambda$ in the corresponding system is large enough or the
interaction range is so small such that $\mathrm {Re}(\mathcal
F_{1,2}(\mu)) <0$.

\item  One bare discrete state coupled with two continuum states with
$v_{11}\neq 0$.

Now, proceeding from previous case,  we gradually turn on $v_{11}\neq 0$ and leave $v_{12}=v_{22}=0$.
Then we  need to examine the zero point of the
determinant of the first $2\times 2$ submatrix, denoted as $\mathbf M_{12}$
\aln{\det \mathbf M_{12}=& ((E-\mu)-\mathcal F_1-\mathcal
F_2)(v_{11}(1-v_{11}G_1(E))) - (v_{11}^2\mathcal F^g_{1}\mathcal
F^{g\ddag}_{1})
\label{eq:M12-1}
\\=&v_{11}(1-v_{11}G_1(E))\left((E-\mu)-\mathcal F_1-\mathcal
F_2 - \frac{(v_{11}^2\mathcal F^g_{1}\mathcal
F^{g\ddag}_{1})}{v_{11}(1-v_{11}G_1(E))}\right).
\label{eq:M12-2}
}
\begin{enumerate}
\item
We first consider $E_\mu<a_1$. We continue to use $E_\mu$ to denote the solution to $M_{11}=(E-\mu)-\mathcal F_1-\mathcal
F_2=0$, meaning that the discrete state is renormalized from the bare mass $\mu$ to $E_\mu$
by turning on $f_1$ and $f_2$. There may also be another bound state
dynamically generated by the continuum-continuum interaction $v_{11}$,
denoted by $E_0$, which arises from the solution to
$1-v_{11}G_1(E)=0$ when $f_1=0$, as discussed in
case~\ref{it:2-cont-1-bound}. We would examine the effect of turning
on $v_{11}$ on the bound state $E_\mu$  and then the effect of
turning on $f_{1,2}$ on the bound state $E_0$.

\begin{itemize}
\item
The simplest case is when $v_{11}>0$,
no bound state is developed by pure
continuum-continuum interaction. Turning on $v_{11}$ will cause the bound state move
upward toward the threshold from $E_\mu$.

This occurs because
at $E=E_\mu$, the last term in \eqref{eq:M12-1} becomes $|v_{11}\mathcal
F^g_{1}(E_\mu)|^2>0$ and $(v_{11}(1-v_{11}G_1(E_\mu)))>0$, while $\mathcal
F_{1,2}(E_\mu)$ are negative. Thus,
turning on $v_{11}$ has the opposite effect of $\mathcal F_{1,2}$
on the discrete state. So, the state moves upward from the previous solution
$E_\mu$ toward the threshold.

\item If $v_{11}<0$, the bound state would always move down from
$E_\mu$.

We first consider the case when $|v_{11}|$ is sufficiently small, such that
 $|v_{11}\mathcal
F^g_{1}(E_\mu)|^2/(v_{11}(1-v_{11}G_1(E_\mu)))<0$. This condition will cause the discrete states
corresponding to $E_\mu$ to move downward.  For larger $|v_{11}|$, there may be a zero point of
$(1-v_{11}G_1(E))$ at $E_0<a_1$, indicating a bound state at $E_0$ that moves  down from the threshold $a_1$,
when $f_1(\omega)=0$. Since at $E_\mu<E_0$,
$\frac{v_{11}^2|\mathcal
F^g_{1}(E_\mu)|^2}{(v_{11}(1-v_{11}G_1(E_\mu)))}<0$,  if
$\mathcal F^g_{1}(E)$ continues to decrease similar to
$G_{1,2}(E)$ for $E<E_0$, the previous result remains valid for large $|v_{11}|$.
In this case, when negative $v_{11}$ is activated and becomes increasingly negative,
the bound state generated from $\mu$ consistently moves down.
\item

For the bound state from $E_0$, switching on  a small interaction
$f_1(\omega)$ will cause the bound state to move upward toward the threshold.

The reasoning is as follows. Since $E_0$ moves down
from the threshold as $v_{11}$ becomes increasingly negative, we have $E_\mu<E_0$ and
$(E_0-\mu)-\mathcal
F_1(E_0)-\mathcal
F_2(E_0)>0$. Then, the negativity of the term $\frac{v_{11}\mathcal F^g_{1}\mathcal
F^{g\ddag}_{1}}{(E-\mu)-\mathcal F_1-\mathcal
F_2}$ causes the bound state corresponding to $E_0$ to shift upward toward the threshold.
Therefore, in this case, turning on $f_1$ appears to activate a repulsive
interaction that decelerates the downward movement of the state at $E_0$ as
$v_{11}$ becomes more attractive. However, turing on $f_2$ will reduce
this deceleration effect, since $-\mathcal F_2$ is positive and
$\frac{v_{11}^2\mathcal F^g_{1}\mathcal
F^{g\ddag}_{1}}{(E-\mu)-\mathcal F_1-\mathcal
F_2}$ will become smaller.

\end{itemize}
\item
Next we look at the case when $a_1<\mu<a_2$,
$f_{1,2}(\omega)$  small
enough, $v_{12}=v_{22}=0$, and
$\mathrm{Re}\mathcal F_1(\mu)<0$, to see the effect of turning on
small $v_{11}$.
By iteration once,  we have an approximation to the solution
\aln{
E_1=\mu+\mathcal F_1(\mu)+\mathcal
F_2(\mu) + \frac{v_{11}\mathcal F^g_{1}(\mu)\mathcal
F^{g\ddag}_{1}(\mu)}{1-v_{11}G_1(\mu)}
}
Expanding to $O(v_{11})$,  we have
\aln{\mathrm{Re} E_1=&\mu+ \mathrm{Re}\mathcal F_1(\mu)+\mathcal
F_2(\mu)+
v_{11}[|\mathrm{P.V.}\mathcal F^g_{1}(\mu)|^2-\pi^2|f_1(\mu)g(\mu)|^2]+O(v_{11}^2)\,,
\label{eq:ReE1}
\\
\mathrm {Im} E_1=&-\pi  \big(|f_1(\mu)|^2+2  v_{11} \mathrm {Re}[\mathcal
F_1^g(\mu) f_1(\mu)g^*(\mu)]\big)+O(v_{11}^2)\,,\label{eq:ImE1}
}
where $\mathrm {P.V.}$ means the principal value part. Thus, the effect of
turning on $v_{11}$ on the mass is determined by the $v_{11}$  term in
Eq. \eqref{eq:ReE1}. If it is positive (negative), it will play an
attractive (repulsive) role.  Whether the width will be
broader or not
depends on the positivity or the negativity of the  second term in the
bracket of Eq.\eqref{eq:ImE1}, respectively.
Using our example form factor, $\mathrm {Re}\mathcal
F_1^g(\mu)\mathrm {Re}( f_1^*(\mu)g(\mu))<0$, a positive (negative) $v_{11}$ causes a
broader (narrower) resonance.
\item When $\mu>a_2$, both $\mathcal
F_{1,2}$ have imaginary parts and  Eqs.~(\ref{eq:ReE1}, \ref{eq:ImE1}) change to
\aln{\mathrm{Re} E_1=&\mu+ \mathrm{Re}\mathcal F_1(\mu)+\mathrm{Re}\mathcal
F_2(\mu)+
v_{11}[|\mathrm{P.V.}\mathcal F^g_{1}(\mu)|^2-\pi^2|f_1(\mu)g(\mu)|^2]+O(v_{11}^2)\,,
\label{eq:ReE2}
\\
\mathrm {Im} E_1=&-\pi  \big(|f_1(\mu)|^2+|f_2(\mu)|^2+2  v_{11} \mathrm {Re}[\mathcal
F_1^g(\mu) f_1(\mu)g^*(\mu)]\big)+O(v_{11}^2)\,.\label{eq:ImE2}
}
The analysis and the result are similar to the previous case.
\end{enumerate}
\item  One bare discrete state coupled with two continuum states with
$v_{12}\neq 0$.

Let us then discuss the effect of nonzero $v_{12}$ and set
$v_{11}=v_{22}=0$. Now the $\mathbf M$ matrix becomes
\begin{align*}
\mathbf M=&\begin{pmatrix}
(E-\mu)-\mathcal F_1(E)-\mathcal F_2(E)& -v_{21}\mathcal
F^g_{2}(E)&- v_{12}\mathcal F^g_{1}(E)
\\
-v^*_{21}\mathcal
F^{g\ddag}_{2}(E)&-|v_{12}|^2G_2(E)&v_{12}
\\
-v^*_{12}\mathcal F^{g\ddag}_{1}(E)&v_{21}&-|v_{12}|^2G_1(E)
\end{pmatrix}
\end{align*}
and
\begin{align}
\det \mathbf M=&|v_{12}|^2\Big(-\big((E-\mu)-\mathcal F_1(E)-\mathcal
F_2(E)\big)\big(1-|v_{12}|^2 G_1(E)G_2(E)\big)
\nonumber \\&+|v_{12}|^2\big(\mathcal
F^g_2(E)\mathcal F^{g\ddag}_2(E)G_1(E)+\mathcal
F^g_1(E)\mathcal F^{g\ddag}_1(E)G_2(E)\big)
\nonumber\\
&+v_{12}^*\mathcal F^{g\ddag}_1(E)\mathcal F^g_2(E)+v_{12}\mathcal
F^{g}_1(E)\mathcal F^{g\ddag}_2(E)\Big).
\label{eq:detM-1D-2C}
\end{align}
This time $v_{12}$ not only appears in $|v_{12}|^2$, but also in
linear terms.

 When $v_{12}=0$, the bare state at $E=\mu$ is renormalized to
$E_\mu<a_1$, which satisfies $(E_\mu-\mu)-\mathcal F_1(E_\mu)-\mathcal
F_2(E_\mu)=0$.
When the coupling $v_{12}$ is turned on, the position of this bound state
will shift. The result depends on the sign of the last line in
Eq.~\eqref{eq:detM-1D-2C} near the bound state. If this term is negative, the effect of turning on
$v_{12}$  will be to pull the bound state downward. Conversely, if it is positive,  a small $|v_{12}|$ will initially
cause the bound state from
$E_\mu$ to move upward. However, as $|v_{12}|$ become sufficiently large, the state will eventually move downward.

The reasoning is as follows.
For $E<a_1$, the $|v_{12}|^2$ term in the second
line of Eq. (\ref{eq:detM-1D-2C}) is always negative and decreases
with $E$. The linear terms of  $v_{12}^{(*)}$ in the third line
takes the form  $2\mathrm
{Re} [v_{12} \mathcal F^{g\ddag}_{1}(E) \mathcal F^g_{2}(E)]$.

We first consider a special case when the two terms in the last line are too small
compared with the second line and can be ignored, for example,
$|\mathcal F_2^g|\ll |\mathcal F_1^g|$ and  $|\mathcal F_2^g|\ll |G_2|$.

\begin{itemize}
\item
Then when $|v_{12}|^2$ is small enough, the factor $1-|v_{12}|^2
G_1(E)G_2(E)>0$ for $E<E_\mu$,
and  the effect of the  purely  negative second line is to push the discrete state downward from
$E_\mu$  as $|v_{12}|$ increases from zero.
\item When $|v_{12}|^2$ becomes large
enough, such that the solution to $1-|v_{12}|^2 G_1(E)G_2(E)=0$ generates $E_b$, which comes down from
the threshold $a_1$, we can expect
$E_\mu<E_b$ since $E_\mu$ is already below the threshold $a_1$.
Given that $G_1(E)G_2(E)>0$ and increases with respect to $E$ below
threshold $a_1$, we still have $1-|v_{12}|^2
G_1(E)G_2(E)>0$ for $E<E_b$.  Thus, the discrete state generated from the bare  state
always goes away from the threshold.
\item We can also examine the effect of the second line on the bound
state generated from $E_b$. Since we have  $(E_b-\mu)-\mathcal F_1(E_b)-\mathcal
F_2(E_b)>0$, the effect of the negative second line is to decelerate the bound state
 from moving down or to pull it toward the threshold $a_1$.

\end{itemize}
Thus, the second line of Eq.\eqref{eq:detM-1D-2C} plays the role
of an effective attractive interaction, dragging the bound state generated from the
bare discrete state downward, while it functions as an effective repulsive interaction for the bound
state arising from the continuum-continuum interaction.

If the last
two terms on the last line can not be ignored, they will add
complexity to the discussion. If the sum of these two term is negative, it will play a similar
role to the second line, whereas if it is positive, it will have the
opposite effect and compete with the second line. In fact, since it is
of order $v_{12}$, it may  contribute more significantly than the second
term for very small $v_{12}$. If this is the case, when $v_{12}$
is activated, the third line will initially dominate the second line. If
both interaction vertices $f_i$ and $g_i$ are real positive
exponential functions, as in the exponential form factor example, the sign
of the last line will correspond to $\mathrm {Re}v_{12}$. A small positive
$\mathrm{Re} v_{12}$ will cause the bare discrete state to move
upward toward the threshold. However, as  $\mathrm{Re} v_{12}$ increases,
the terms in the second line will dominate, dragging
the discrete state down from $E_\mu$ and decelerating the one from $E_b$
from descending.  There is also the possibility that the third
line is sufficiently large such that the bound state from $E_b$ collides with
the bound state generated from $E_\mu$ as $|v_{12}|$ increases,
and then they may separate again into two bound state again, one moving downward and the other moving upward.

\end{enumerate}
In more complicated cases, the results may be intricate, and may
not present a simple picture. The previous cases serve as examples
for the analyzing the effects of the different interaction in
various situations and qualitatively understanding the behavior of the
pole positions.

\section{Conclusion\label{sect:conclude}}

 This paper presents several improvements to the Friedrichs
model, aiming to provide a more comprehensive description of coupled
channel scattering in real-world scenarios. Firstly, we investigate
situations involving multiple discrete states and continuum states,
focusing on the general interaction between these discrete states and
the continuum states. Secondly, we consider the inclusion of
continuum-continuum interactions, employing a more general
separable interaction that is independent of the interaction between
the discrete states and the continuum states. Notably, this extended
model remains exactly solvable. Thirdly, we address scenarios where
the square integrable interaction between the continuum states takes a
non-separable form, rendering it non-solvable. However, we propose an
approach to approximate this potential by expanding it in terms of a
chosen basis set, effectively expressing it as a truncated series of
separable potentials. Consequently, at a finite order, this potential
becomes solvable. To simplify the analysis, we also suggest utilizing the
same basis set for expanding both the discrete-continuum interaction
and the continuum-continuum potential.  A few simple examples are
discussed to analyse the behaviors of the masses of the discrete
states when different interaction are turned on, which may be helpful in
qualitatively understanding the spectrum in the coupled channel system.
A few interesting results may also be useful for the systems where the
couplings between states can be tuned such as the cold atom systems.

This discussion establishes a theoretical foundation for the
application of the Friedrichs model in various contexts, including
hadron physics and other areas involving coupled channel scattering
and intermediate resonances. To utilize the model effectively, one
must first model the interaction between the discrete-continuum and
continuum-continuum components. Subsequently, the continuum-continuum
potential can be approximated using a series of separable potentials,
enabling resonance searches or $S$-matrix calculations. An
advantageous aspect of this model is the automatic preservation of
unitarity in the $S$-matrix, while avoiding the presence of spurious
poles on the first Riemann sheet. In contrast, the conventional
$K$-matrix parameterization lacks control over spurious poles on the
physical sheet.

However, a remaining challenge lies in determining the
continuum-continuum interactions in a reasonable manner. Further
research is required to develop suitable approaches for obtaining
these interaction terms in a manner that meets the physical
expectations and provides reliable results in various real world applications.

\begin{acknowledgments}
 This work is supported by China National Natural Science Foundation
under contract No. 11975075, No. 12375132, No. 12375078, No. 11575177, No.11947301, and No.12335002.
This work is also supported by “the Fundamental Research Funds for the Central Universities”.
\end{acknowledgments}

\appendix
\section{The detailed derivation of the normalization and the $S$
matrix in section \ref{sect:sep-cont}}\label{sect:sep-cont:app}

The normalization of the continuum state using the coefficients in
(\ref{eq:discret-eq-cont}) and (\ref{eq:cont-eq-cont})  can be
calculated as follows,
\aln{
\langle
\Psi^\pm_m(E)|\Psi^\pm_n(E')\rangle=&\sum_{i=1}^D\alpha_{im}^*(E)\alpha_{in}(E')+\delta_{mn}\delta(E-E')
\nonumber\\&
+\frac
1{E-E'\mp i0}\Big(\sum_{j=1}^D \frac{A^*_{jm}(E)
f^*_{jn}(E')}{E-M_j}+\sum_{n'=1}^C v^*_{nn'}B^*_{n'm}(E)g^*_n(E')\Big)
\nonumber\\&
+\frac
1{E'-E\pm i0}\Big(\sum_{j=1}^D \frac{A_{jm}(E')
f_{jn}(E)}{E'-M_j}
+\sum_{n'=1}^C v_{nn'}B_{n'm}(E')g_n(E)\Big)
\nonumber\\&+\sum_{m'=1}^C\int_{a_{m'}} d\omega\underbrace{\frac
1{E-\omega\mp i0}\frac
1{E'-\omega\pm i0}}_{\frac1{E'-E\pm i0}\big(\frac
1{E-\omega\mp i0}-\frac
1{E'-\omega\pm i0}\big) }\Big(\sum_{j=1}^D \frac{A^*_{jm}(E)
f^*_{jm'}(\omega)}{E-M_j}+\sum_{n'=1}^C v^*_{m'n'}B^*_{n'm}(E)g^*_{m'}(\omega)\Big)
\nonumber\\&\times\Big(\sum_{j'=1}^D \frac{A_{j'n}(E')
f_{j'm'}(\omega)}{E'-M_{j'}}+\sum_{n''=1}^C v_{m'n''}B_{n''n}(E')g_{m'}(\omega)\Big).
\label{eq:Norm-cont}
}
Notice that we have omitted the $\pm$ superscripts in the coefficients
$\alpha_{im}$, $A_{jm}$ and $B_{mn}$ since they all have the same superscript of $\pm$.
Using the definitions in Eq.~(\ref{eq:def-M}), Eqs.~(\ref{eq:MY-components-1}) and
(\ref{eq:MY-components-2}), the last two lines can be reduced as
\als{
\frac1{E'-E\pm
i0}\Big[&\sum_{j,j'=1}^D{
\alpha^{\pm*}_{jm}(E)}\big(\delta_{jj'}(E-E'){
-V^{\pm*}_{j'j}(E)}+{
V^{\pm}_{jj'}(E')}\big){\alpha^\pm_{j'n}(E')}
\\&+\sum_{j=1}^D\sum_{n''=1}^C{ \alpha^{\pm*}_{jm}(E)}\big({ -\sum_{m'}V^{\pm
BA*}_{m'j}(E)v_{m'n''}}+ {V^{\pm AB}_{jn''}(E')}\big)
{
B^\pm_{n''n}(E')}
\\&+\sum_{j'=1}^D\sum_{n'=1}^C{ B^{\pm*}_{n'm}}\big({ -V^{\pm
AB*}_{j'n'}(E)}+\sum_{m'=1}^Cv^*_{m'n'}{ V^{\pm
BA}_{j'm'}}\big){ \alpha^\pm_{j'n}(E')}
\\&+\sum_{n',n'',m'=1}^C{ B^{\pm*}_{n'm}(E)}\big({ -V^{\pm
B*}_{m'n'}(E)v_{m'n''}}+v^*_{m'n'}{ V^{\pm
B}_{m'n''}(E')}\big){ B^\pm_{n''n}(E')}\Big]
\\&\hspace{-3cm}=-\sum_{j=1}^D\alpha^{\pm*}_{jm}(E)\alpha^\pm_{jn}(E)+ \frac1{E'-E\pm
i0}\Big[\sum_{j'=1}^D({
-f_{j'm}(E)})\alpha^\pm_{j'n}(E')+\alpha^{\pm*}_{j'm}(E){
f^*_{j'n}(E')})
\\&+\sum_{n',m'=1}^C({-\delta_{mm'}g_m(E)}
v_{m'n'}B^\pm_{n'n}(E')+B^{\pm*}_{n'm}(E)v^*_{m'n'}{
g^*_{m'}(E')\delta_{m'n})}\Big].
}
Thus, they cancel with the terms in Eq.~(\ref{eq:Norm-cont}) except
for the $\delta(E-E')$ term.
The $S$-matrix can also be obtained by
\aln{\langle \Psi^-_m(E)|\Psi^+_n(E')\rangle=\sum_{i=1}^D
&\sum_{i=1}^D\alpha_{im}^{-*}(E)\alpha^+_{in}(E')+\delta_{mn}\delta(E-E')
\nonumber\\&
+(\frac
1{E-E'- i0}{-2\pi i\delta(E-E')})\Big(\sum_{j=1}^D \frac{A^{-*}_{jm}(E)
f^*_{jn}(E')}{E-M_j}+\sum_{n'=1}^C v^*_{nn'}B^{-*}_{n'm}(E)g^*_n(E')\Big)
\nonumber\\&
+\frac
1{E'-E+ i0}\Big(\sum_{j=1}^D \frac{A^+_{jm}(E')
f_{jn}(E)}{E'-M_j}
+\sum_{n'=1}^C v_{nn'}B^+_{n'm}(E')g_n(E)\Big)
\nonumber\\&+\sum_{m'=1}^C\int_{a_{m'}} d\omega\underbrace{\frac
1{E-\omega+ i0}\frac
1{E'-\omega+ i0}}_{\frac1{E'-E+ i0}\big(\frac
1{E-\omega+ i0}-\frac
1{E'-\omega+ i0}\big)}\Big(\sum_{j=1}^D \frac{A^{-*}_{jm}(E)
f^*_{jm'}(\omega)}{E-M_j}+\sum_{n'=1}^C v^*_{m'n'}B^{-*}_{n'm}(E)g^*_{m'}(\omega)\Big)
\nonumber\\&\times\Big(\sum_{j'=1}^D \frac{A^+_{j'n}(E')
f_{j'm'}(\omega)}{E'-M_{j'}}+\sum_{n''=1}^C v_{m'n''}B^+_{n''n}(E')g_{m'}(\omega)\Big).
\label{eq:S-matrix-couplednew}
}
We have used $\frac
1{E-\omega+ i0}\frac
1{E'-\omega+ i0}=\frac1{E'-E+ i0}\big(\frac
1{E-\omega- i0}-\frac
1{E'-\omega+ i0}\big)-2\pi i\delta(E-\omega) \frac
1{E'-\omega+ i0}=\frac1{E'-E+ i0}\big(\frac
1{E-\omega+ i0}-\frac
1{E'-\omega+ i0}\big)$. Since $\delta(E'-E)\big(\frac
1{E-\omega+ i0}-\frac
1{E'-\omega+ i0}\big)=0$, the $i0$ in the first factor does not have
any effect.
The last two lines can be reduced to
\als{
\frac1{E'-E+
i0}\Big[&\sum_{j,j'=1}^D{
\alpha^{-*}_{jm}(E)}\big(\delta_{jj'}(E-E'){
-V^{-*}_{j'j}(E)}+{
V^{+}_{jj'}(E')}\big){\alpha^+_{j'n}(E')}
\\&+\sum_{j=1}^D\sum_{n''=1}^C{ \alpha^{-*}_{jm}(E)}\big({
-\sum_{m'}V^{-
BA*}_{m'j}(E)v_{m'n''}}+ { V^{+ AB}_{jn''}(E')}\big)
{
B^+_{n''n}(E')}
\\&+\sum_{j'=1}^D\sum_{n'=1}^C{ B^{-*}_{n'm}}\big({ -V^{-
AB*}_{j'n'}(E)}+\sum_{m'=1}^Cv^*_{m'n'}{ V^{+
BA}_{j'm'}}\big){ \alpha^+_{j'n}(E')}
\\&+\sum_{n',n'',m'=1}^C{ B^{-*}_{n'm}(E)}\big({ -V^{-
B*}_{m'n'}(E)v_{m'n''}}+v^*_{m'n'}{ V^{+
B}_{m'n''}(E')}\big){ B^+_{n''n}(E')}\Big]
\\&\hspace{-3cm}=-\sum_{j=1}^D\alpha^{-*}_{jm}(E)\alpha^+_{jn}(E)+
\frac1{E'-E+
i0}\Big[\sum_{j'=1}^D({
-f_{j'm}(E)})\alpha^+_{j'n}(E')+\alpha^{-*}_{j'm}(E){
f^*_{j'n}(E')})
\\&+\sum_{n',m'=1}^C({-\delta_{mm'}g_m(E)}
v_{m'n'}B^+_{n'n}(E')+B^{-*}_{n'm}(E)v^*_{m'n'}{
g^*_{n}(E')\delta_{m'n})}\Big].
}
These terms cancel with the other terms except the terms with $\delta(E-E')$.
Notice that for $E=E'$ and is real, the final sum inside the square bracket will be
zero,
\als{
\Big[&\sum_{j'=1}^D({
-f_{j'm}(E)}\alpha^+_{j'n}(E)+\alpha^{-*}_{j'm}(E){
f^*_{j'n}(E)})
\\&+\sum_{n',m'=1}^C({-\delta_{mm'}g_m(E)}
v_{m'n'}B^+_{n'n}(E)+B^{-*}_{n'm}(E)v^*_{m'n'}{
g^*_{m'}(E)\delta_{m'n})}\Big]=0,
}
which can be derived directly from Eqs.~(\ref{eq:MY-components-1}) and
(\ref{eq:MY-components-2}). Then the $S$ matrix can be derived:
\als{
S_{mn}(E,E')=&\delta(E-E')-2\pi i\delta(E-E')\Big(\sum_{j=1}^D \frac{A^{-*}_{jm}(E)
f^*_{jn}(E')}{E-M_j}+\sum_{n'=1}^C v^*_{nn'}B^{-*}_{n'm}(E)g^*_n(E')\Big)
\\
=&\delta(E-E')-\pi i\delta(E-E')\Big(\sum_{j=1}^D\big({  \alpha^{-*}_{jm}(E)
f^*_{jn}(E')}+{ f_{j'm}(E)\alpha^+_{j'n}(E')}\big)
\\&+\sum_{n'=1}^C
\big({ v^*_{nn'}B^{-*}_{n'm}(E)g^*_n(E')}+
{ v_{mn'}B^{+}_{n'n}(E)g_m(E')}\big)\Big)\,.
}
From the definition of $\mathbf Y$ and $\mathbf F$ in
Eq.~(\ref{eq:Mat-Y-F}) and solving
$\mathbf Y$
from Eq.(\ref{eq:MY-F}), we can reformulate
the previous equation using the matrices and obtain Eq.~(\ref{eq:S-FM-1F}).

\section{Solving the approximate contact potential\label{sect:solu-Approx-4p}}
This section we provide the details for solving the eigenstate
problem for the Hamiltonian in section \ref{sect:Approx-Sep}.
After approximating the general contact potential as the sum of the separable
potentials, the approximated Hamiltonian for multiple continuum states and discrete
states is shown in  Eq.(\ref{eq:H-Approx-Sep}) which is copied here for completeness
\begin{align}
H
=&\sum_{i=1}^D M_i|i\rangle\langle
i|+\sum_{n=1}^C \int_{a_n}^\infty \mathrm d \omega
\,\omega|\omega;n\rangle\langle \omega;n|
\nonumber \\ &+\sum_{m,n=1}^C v_{mn,\rho\delta}\Big(\int_{a_m}^\infty\mathrm d
\omega'
{ \tilde g_{m\rho}(\omega')}|\omega';m\rangle\Big)\Big(\int_{a_n}^\infty\mathrm d \omega
{ \tilde g^*_{n\delta}(\omega)}\langle \omega;n|\Big)
\nonumber \\ &+\sum_{j=1}^D\sum_{n=1}^C \left[ |j\rangle\Big(\int_{a_n}^\infty\mathrm d \omega
{ f^*_{jn}(\omega)}\langle \omega;n|\Big)+\Big(\int_{a_n}^\infty\mathrm d \omega
{ f_{jn} (\omega)}|\omega;n\rangle \Big)
\langle j|
\right]\,.
\end{align}
The general eigenstate for this eigenvalue problem can be expanded using
the bare discrete states and the bare continuum states
\begin{align*}
|\Psi(E)\rangle
=&\sum_{i=1}^D
\alpha_i(E)|i\rangle+\sum_{n=1}^C\int_{a_n} \mathrm d\omega
\psi_{n}(E,\omega)|\omega;n\rangle\,.
\end{align*}
The proceeding derivation goes in parallel with the process in section
\ref{sect:sep-cont}.
With this ansatz, the eigenvalue problem can be reduced to the
following equations
\begin{align*}
(&M_j-E)\alpha_j(E)+A_j(E)=
0,\quad
j=1,\dots,D
\\
&\sum_{j=1}^D\alpha_j
(E)f_{jn}(\omega)+(\omega-E)\psi_{n}(E,\omega)+\sum_{m=1}^Cv_{nm,\rho\delta}{
\psi_{m\delta}(E)}
\tilde g_{n\rho}(\omega)
=0,
\quad n=1,\dots,C, \quad\mathrm{and\ }\omega>a_n
\end{align*}
where we have defined
\aln{A_j(E)
=\sum_{n=1}^C\int_{a_n}^\infty\mathrm
d\omega\, f^*_{jn}(\omega)\psi_n(E,\omega),
\quad \psi_{n\delta}(E)=\int d\omega \psi_{n}(E,\omega)\tilde
g^*_{n\delta}(\omega).
\label{eq:A-psi-def}
}
There are $C$ continuum eigenstate solutions and $|\Psi(E)\rangle$, $\alpha_i$, $\psi_{n\delta}$
and $A_{j}(E)$
need another index $m$ to denote different continuum solutions, i.e. $|\Psi_m(E)\rangle$, $\alpha_{im}(E)$, $\psi_{nm\delta}(E)$, and
$A_{jm}(E)$. Similar to section \ref{sect:sep-cont}, we require that
$|\Psi_m(E)\rangle$ tends to $|E,m\rangle$ as the interactions are turned
off and consider the $C$ continuum solutions for $E>a_C$.
Then the above equations can be reduced to
\begin{align}
\alpha^\pm_{jm}(E)=&\frac1{E-M_j}A^\pm_{jm}(E)\,,
\label{eq:discret-VExp-cont}
\\
\psi^\pm_{nm}(E,\omega)
=&\delta_{nm}\delta(E-\omega)+
\sum_{j=1}^D \frac{\alpha^\pm_{jm}(E)f_{jn}(\omega)}{E-\omega\pm i0}
+\sum_{n'=1}^C
v_{nn',\delta'\rho}\frac{\psi^\pm_{n'm\rho}(E)\tilde
g_{n\delta'}(\omega)}{E-\omega\pm i0}.
\label{eq:cont-VExp-cont}
\end{align}

By applying the operation $\sum_n
\int_{a_n} d\omega f^*_{jn}(\omega)\times$
and  the operations $\sum_{n\delta}v^*_{nm',\rho'\delta}\int_{a_n} d\omega\tilde
g^*_{n\delta}(\omega)\times$ on (\ref{eq:cont-VExp-cont})
respectively,
 we obtain
\aln{
0=&-f^*_{jm}(E)+\sum_{j'=1}^D\Big((E-M_{j'})\delta_{jj'}-
\sum_{n=1}^C\int_{a_n}d\omega\frac{f_{j'n}(\omega)f^*_{jn}(\omega)}{E-\omega\pm
i0}\Big)\frac{A^\pm_{j'm}(E)}{E-M_{j'}}
-\sum_{n',n=1}^C
v_{nn',\delta'\rho}\psi^\pm_{n'm\rho}(E)\int_{a_n}d\omega\frac{f^*_{jn}(\omega)\tilde
g_{n\delta'}(\omega)}{E-\omega\pm i0}\,,
\label{eq:disc-eq-A}\\
0=
&-v^*_{mm',\delta\rho'}\tilde g^*_{m\delta}(E)-\sum_{j=1}^D {
\alpha^\pm_{jm}(E)}\sum_{n=1}^Cv^*_{nm',\delta\rho'}
\int_{a_n} d\omega \frac{{ f_{jn}(\omega)}\tilde g_{n\delta}^*(\omega)}{E-\omega\pm i0}
\nonumber\\&\quad\quad
+
\sum_{n'}^C\psi^\pm_{n'm\rho}(E)
\bigg[v^*_{n'm',\rho\rho'}-\sum_{n=1}^Cv^*_{nm',\delta\rho'}v_{nn',\delta'\rho}\int_{a_n} d\omega \frac{\tilde
g_{n\delta'}(\omega)\tilde g_{n\delta}^*(\omega)}{E-\omega\pm
i0}\bigg].
\label{eq:cont-eq-psi}
}
These two equations correspond to previous Eq.~(\ref{eq:cont-eq-f})
and (\ref{eq:cont-eq-g}). Notice that the differences are
Greek letters and the sums here.
Similar to the vectors and matrices defined in Eqs.(\ref{eq:Mat-Y-F})-(\ref{eq:def-M}), we define
\aln{
&(\tilde {\mathbf F}_m)_{m'\rho'}=v^*_{mm',\delta\rho'}\tilde
g^*_{m\delta}(E)\,,\quad (\tilde {\mathbf F}_m)_j=f^*_{jm}(E)\,,\quad
m=1,\dots, C; j=1,\dots,D; \delta=1,2,\dots, N
\nonumber
\\ &(\tilde {\mathbf Y}_m(E))_{j}=\alpha_{jm}(E) \,,\quad(\tilde
{\mathbf Y}_m)_{n'\rho}=\psi^\pm_{n'm\rho}(E)\,, \quad j=1,\dots,D;
m,n'=1,\dots,C;\rho=1,2,\dots, N
\label{eq:tilde-Y}
\\&\tilde {\mathbf V}^{BB}_{m'\rho',n'\rho}=v^*_{n'm',\rho\rho'}-\sum_{n=1}^Cv^*_{nm',\delta\rho'}v_{nn',\delta'\rho}\int_{a_n} d\omega \frac{\tilde
g_{n\delta'}(\omega)\tilde g_{n\delta}^*(\omega)}{E-\omega\pm
i0},\quad m',n'=1,2,\dots,C; \rho',\rho=1,2,\dots, N
\nonumber
\\&\tilde {\mathbf V}^{AA}_{jj'}=(E-M_{j'})\delta_{jj'}-
\sum_{n=1}^C\int_{a_n}d\omega\frac{f_{j'n}(\omega)f^*_{jn}(\omega)}{E-\omega\pm
i0},\quad j,j'=1,2,\dots, D
\nonumber\\&
\tilde {\mathbf V}^{AB}_{j,n'\rho}=-\sum_{n=1}^Cv_{nn',\delta'\rho}\int_{a_n}d\omega\frac{f^*_{jn}(\omega)\tilde
g_{n\delta'}(\omega)}{E-\omega\pm i0}, \quad j=1,\dots,D; n'=1,\dots,
C;\rho=1,2,\dots, N
\nonumber\\&
\tilde {\mathbf V}^{BA}_{m'\rho',j}=-\sum_{n=1}^Cv^*_{nm',\delta\rho'}
\int_{a_n} d\omega \frac{{ f_{jn}(\omega)}\tilde g_{n\delta}^*(\omega)}{E-\omega\pm i0}, \quad j=1,\dots,D; m'=1,\dots,
C;\rho'=1,2,\dots, N
\nonumber\\
&\tilde{\mathbf M}_{IJ}=\begin{pmatrix} \tilde {\mathbf V}^{AA}(E)&\tilde
{\mathbf V}^{AB}(E)\\
\tilde {\mathbf V}^{BA}(E)&\tilde {\mathbf V}^{BB}(E)\end{pmatrix}_{IJ}.\quad
IJ=1,\dots,D, D+1,\dots, D+NC\,.
}
We still have $\tilde {\bf M}^{+\dagger} =\tilde {\bf M}^-$. With these
matrices, Eqs. (\ref{eq:disc-eq-A}) and (\ref{eq:cont-eq-psi}) can be expressed as
\[ \tilde {\bf M}\cdot \tilde {\bf Y}_m=\tilde {\bf F}_m\]
or
\aln{
\sum_{j=1}^D\tilde{\bf V}^{AA}_{ij}(E) \alpha_{jm}(E)+\sum_{n'=1}^C\tilde{\bf V}^{AB}_{i,n'\rho}(E)
\psi_{n'm\rho}(E)=&f^*_{im}(E),
\\
\sum_{j=1}^D\tilde{\bf V}^{BA}_{n\delta,j}(E) \alpha_{jm}(E)+\sum_{n'=1}^C\tilde{\bf V}^{BB}_{n\delta,n'\rho}(E)
\psi_{n'm\rho}(E)=&v^*_{nn',\delta\rho}(E)\tilde g^*_{m\rho}(E).
}
As before $\tilde{\mathbf M}$ is still independent of $m$, but
$\tilde {\mathbf F}_m$ depends on $m$. If there are infinite number of bases,
the matrix $\tilde {\mathbf M}$ and vector $\tilde{\bf F}_m$ and
$\tilde {\mathbf Y}$ are infinite dimensional. Now we have supposed
that the bases
chosen are well enough, and have made a truncation to a finite order
$N$ of the expansion of the
potential $V_{nn'}$ i.e.  $v_{nn',\delta\rho}=0$ for $\delta,\rho > N$.
Then $\tilde {\bf M}$ is a   $(D+N C)\times (D+ NC)$ matrix.
In general, the matrix $\tilde {\bf M}$ is non-degenerate for $E>a_m$,
and $\tilde{\bf Y}_m$ can be solved,
\aln{\tilde{\bf Y}_m(E)=\tilde{\bf
M}^{-1}\cdot \tilde {\bf F}_m.
\label{eq:tildY-solu}}
With all the
$\psi^\pm_{nm\rho}(E)$ and $\alpha^\pm_{jm}(E)$ at hand,  the
approximate continuum solutions are
solved as
\begin{align}
|\Psi^\pm_m(E)\rangle
=&\sum_{i=1}^D
\alpha^\pm_{im}(E)|i\rangle+|E;m\rangle+
\sum_{n=1}^C\int_{a_n} \mathrm d\omega\frac{1}{E-\omega\pm i0}
\Big(\sum_{j=1}^D \alpha^\pm_{jm}(E)f_{jn}(\omega)
+\sum_{n'=1}^C
v_{nn',\delta'\rho}\psi^\pm_{n'm\rho}(E)\tilde
g_{n\delta'}(\omega)
\Big)|\omega;n\rangle\,\nonumber
\\
=&|E;m\rangle+\sum_{j=1}^D
\alpha^\pm_{jm}(E)\Big(|j\rangle+
\sum_{n=1}^C\int_{a_n} \mathrm d\omega\frac{f_{jn}(\omega)}{E-\omega\pm i0}
|\omega;n\rangle\Big)
+\Big( \sum_{n,n'=1}^Cv_{nn',\delta'\rho}\psi^\pm_{n'm\rho}(E)\int_{a_n} \mathrm d\omega\frac{\tilde
g_{n\delta'}(\omega)}{E-\omega\pm i0}
|\omega;n\rangle\Big).
\end{align}
It can be checked that the normalization  is $\langle
\Psi_m^\pm(E)|\Psi_n^\pm(E')\rangle=\delta_{mn}\delta(E-E')$.
The $S$-matrix can be obtained,
\aln{
\langle
\Psi_m^-(E)|\Psi_n^+(E')\rangle=&\delta_{mn}\delta(E-E')
-2\pi i\delta(E-E')\Big(\tilde{\bf F}^\dagger_m \cdot (\tilde{\bf
M}^+)^{-1}\cdot \tilde{\bf F}_m
\Big).
}
For  discrete eigenvalues and discrete eigenstates, there will not be
the delta function in Eq.~(\ref{eq:cont-VExp-cont}), and we have
equations,
\[\tilde {\bf M} \cdot \tilde {\bf Y}=0,\]
where $\tilde {\bf Y}$ is defined similar to Eq. (\ref{eq:tilde-Y}) without
subindex $m$.
The generalized energy eigenvalues for the discrete state can be
obtained from the $\det \tilde {\mathbf M}(E)=0$ and the eigenvector
$\tilde {\bf Y}$ can be obtained with proper normalization chosen as in
previous section. Then the $i$-th discrete state can be expressed as
\begin{align}
|\Psi^{(i)}(E_i)\rangle
=&\sum_{i=1}^D
\alpha^{(i)}_{j}(E_i)\Big(|j\rangle+
\int_{a_n} \mathrm d\omega\frac{f_{jn}(\omega)}{E_i-\omega}
|\omega;n\rangle\Big)
+\sum_{n'=1}^C
v_{nn',\delta'\rho}\psi^{(i)}_{n'm\rho}(E_i)\int_{a_n} \mathrm d\omega\frac{\tilde
g_{n\delta'}(\omega)
}{E_i-\omega}
|\omega;n\rangle\,,
\\
&\mathrm {with\ } \tilde {\mathbf Y}^{(i)}(E_i)\cdot \tilde {\mathbf M}'(E_i)\cdot \tilde
{\mathbf Y}^{(i)*}(E_i)=1
\nonumber\end{align}
where $\tilde {\mathbf M}'(E)$ is the derivative of the matrix w.r.t.
$E_i$ and  the integral contour needs to be deformed for $E_i$ on
unphysical sheets as before.

\section{Solving the eigenvalue problem with both approximate vertex
functions and contact interactions~\label{sect:solu-Approx-3p}}
This section serves to solve the eigenstates for the Hamiltonian in
Eq.(\ref{eq:H-Approx-Sep-f}) where both the contact potential and
the vertex are expanded using the function bases $\tilde g_{n\delta}$,
which we reproduce here for completeness
\begin{align}
H=&\sum_{i=1}^D M_i|i\rangle\langle
i|+\sum_{n=1}^C \int_{a_n}^\infty \mathrm d \omega
\,\omega|\omega;n\rangle\langle \omega;n|
\nonumber\\ &+\sum_{m,n=1}^C v_{mn,\rho\delta}\Big(\int_{a_m}^\infty\mathrm d
\omega'
{ \tilde g_{m\rho}(\omega')}|\omega';m\rangle\Big)\Big(\int_{a_n}^\infty\mathrm d \omega
{ \tilde g^*_{n\delta}(\omega)}\langle \omega;n|\Big)
\nonumber\\ &+\sum_{j=1}^D\sum_{n=1}^C \left[f^*_{jn\delta} |j\rangle\Big(\int_{a_n}^\infty\mathrm d \omega
{ \tilde g^*_{n\delta}(\omega)}\langle \omega;n|\Big)+{ f_{jn\delta }} \Big(\int_{a_n}^\infty\mathrm d \omega
\tilde g_{n\delta}(\omega)|\omega;n\rangle \Big)
\langle j|
\right].
\end{align}
This case is more like the cases discussed
in~\cite{Xiao:2016mon,Sekihara:2014kya}, where the same form factor
comes with the continuum both in the discrete-continuum and
continuum-continuum interaction.
Using the eigenstate ansatz
\begin{align}
|\Psi(E)\rangle
=&\sum_{i=1}^D
\alpha_{i}(E)|i\rangle+\int_{a_n} \mathrm d\omega
\psi_{n}(E,\omega)|\omega;n\rangle
\nonumber\\
=&\sum_{i=1}^D
\alpha_i(E)|i\rangle+\sum_{n=1}^C\psi_{n\delta}(E)\int_{a_n} \mathrm d\omega
\tilde g_{n\delta}(\omega)|\omega;n\rangle\,,
\end{align}
and proceeding in solving the eigenvalue problem similarly to the previous
section, one finds that $A_j$ in Eq.~(\ref{eq:A-psi-def}) becomes
\[A_j(E)
=\sum_{n=1}^C
\, f^*_{jn\delta}\psi_{n\delta}(E).
\]
For the $m$-th continuum solution
$|\Psi_m(E)\rangle$, the coefficients $\alpha_i$, $\psi_n$ and
$\psi_{n\delta}$ take another subindex $m$ and
Eq.~(\ref{eq:discret-VExp-cont}) and (\ref{eq:cont-VExp-cont})
becomes
\begin{align}
\alpha^\pm_{jm}(E)=&\frac1{E-M_j}A^\pm_{jm}(E)=\frac1{E-M_j}\sum_{n'=1}^C
f^*_{jn'\rho}\psi^\pm_{n'm\rho}(E),
\\\psi^\pm_{nm}(E,\omega)
=&\delta_{nm}\delta(E-\omega)+
\sum_{n'=1}^C\psi^\pm_{n'm\rho}(E)
V_{n\delta',n'\rho}(E)\frac{\tilde
g_{n\delta'}(\omega)}{E-\omega\pm i0},
\label{eq:psi-alpha}
\end{align}
where $V_{n\delta',n'\rho}(E)$ is defined as the matrix elements of
a $NC\times NC$ matrix
$\mathbf V$
\[(\mathbf V{ (E)})_{n\delta',n'\rho}= \sum_{j=1}^D\frac{f_{jn\delta'}f^*_{jn'\rho}}{E-M_j}
+
v_{nn',\delta'\rho},\]
which is supposed to be non-degenerate for general $E$. Multiplying $\tilde g_{n\delta}^*(\omega)$ to above equation
(\ref{eq:psi-alpha})
and integrating w.r.t $\omega$, we get
\begin{align}
\delta_{nm}\tilde g^*_{n\delta}(E)=
\sum_{n'=1}^C\psi^\pm_{n'm\rho}(E)
\bigg[\delta_{nn'}\delta_{\rho\delta}-V_{n \delta',n'\rho}(E)\int_{a_n} d\omega \frac{\tilde
g_{n\delta'}(\omega)\tilde g_{n\delta}^*(\omega)}{E-\omega\pm
i0}\bigg].
\label{eq:cont-expand-eq}
\end{align}
Define $NC\times NC$ matrix $\tilde {\mathbf M}$ and $NC$ dimensional vector $\tilde
{\bf F}_m$, $\tilde {\bf Y}_m$ as
\aln{
&\tilde {\mathbf M}_{n\delta,n'\rho}=\delta_{nn'}\delta_{\rho\delta}-V_{n \delta',n'\rho}(E)\int_{a_n} d\omega \frac{\tilde
g_{n\delta'}(\omega)\tilde g_{n\delta}^*(\omega)}{E-\omega\pm
i0},
\label{eq:M-Ext}\\
&(\tilde {\mathbf F}_m)_{n\delta}=\delta_{nm}\tilde
g^*_{n\delta}(E)\,,\quad (\tilde {\mathbf Y}_m)_{n'\rho}=\psi^\pm_{n'm\rho}(E).
\label{eq:tilde-F-Y}}
Then Eq.(\ref{eq:cont-expand-eq}) can be expressed as $\tilde {\mathbf
F}_m=\tilde{ \mathbf M}\cdot \tilde{\mathbf Y}_m$.
Notice that $\tilde{\mathbf M}$ is independent of the $m$-th solution, but
$\tilde {\mathbf F}_m$ depends on $m$. To see the real analyticity of
$\det \tilde {\mathbf M}$, we define
\aln{(\tilde{\bf W}^\pm)_{n\delta,n'\rho}=(\tilde {\bf M}^\pm\cdot \mathbf V^{-1})_{n\delta,n'\rho}={\mathbf V}^{-1}_{n\delta,n'\rho}-\delta_{nn'}\int_{a_n} d\omega \frac{\tilde
g_{n\rho}(\omega)\tilde g_{n\delta}^*(\omega)}{E-\omega\pm
i0},\label{eq:tilde-W}} 
and then  we have $\tilde {\mathbf M}^\pm=\tilde{\mathbf W}^\pm\cdot \mathbf V$. Since
$(\tilde {\mathbf W}^\pm(E))^\dagger =\tilde {\mathbf W}^\mp(E)$
and $\mathbf V(E)$ is hermitian
for real $E$,
$\det\tilde {\mathbf M}^{\pm \dagger }(E)=\det\tilde {\mathbf
M}^{\mp}(E)$. {So the analytically continued $\det \tilde {\mathbf
M}(E)$ with $\det\tilde {\mathbf M}^{+}(E)$ and $\det\tilde {\mathbf M}^{-}(E)$
on the upper and lower edge of the cut above the threshold satisfies
the Schwartz reflection relation, $\det\tilde {\mathbf M}^{*}(E)=\det\tilde {\mathbf
M}(E^*)$.}

Then, in general, the matrix $\tilde {\bf M}$ is non-degenerate for $E>a_m$,
and $\tilde{\bf Y}_m$ can be solved,
\aln{\psi^\pm_{nm\rho}(E)=(\tilde{\bf Y}_m(E))_{n\rho}=(\tilde{\bf
M}^{-1}\cdot \tilde {\bf F}_m)_{n\rho}=(\tilde{\bf
M}^{-1})_{n\rho,m\delta} \tilde g_{m\delta}(E).
\label{eq:psi-tilde}}
With all the
$\psi^\pm_{nm\rho}(E)$ for $\rho\le N$ at hand,  all the
$\alpha^\pm_{jm}(E)$ and $\psi^\pm_{nm}(E,\omega)$  can also
be obtained.
Then, the continuum state can be approximated as
\begin{align}
|\Psi^\pm_m(E)\rangle
=&\sum_{i=1}^D
\alpha^\pm_{im}(E)|i\rangle+|E,m\rangle+\sum_{n=1}^C\sum_{n'=1}^C\psi^\pm_{n'm\rho}(E)
V_{n\delta',n'\rho}(E)\int_{a_n} \mathrm d\omega
\frac{\tilde
g_{n\delta'}(\omega)}{E-\omega\pm i0}
|\omega;n\rangle.
\label{eq:expand-Ext-sol}
\end{align}
It can be checked that the normalization  is $\langle
\Psi_m^\pm(E)|\Psi_n^\pm(E')\rangle=\delta_{mn}\delta(E-E')$.
The $S$-matrix can be obtained
\aln{
\langle
\Psi_m^-(E)|\Psi_n^+(E')\rangle=&\delta_{mn}\delta(E-E')-2\pi
i\delta(E-E')\sum_{n=1}^C\sum_{n'=1}^C\psi^{-*}_{n'm\rho}(E')
V^*_{n\delta',n'\rho}(E')
\tilde
g^*_{n\delta'}(E)
\nonumber\\=&\delta_{mn}\delta(E-E')-2\pi
i\delta(E-E')(\tilde {\mathbf F}_m^\dagger\cdot (\tilde {\mathbf W}^{+})^{-1}
\cdot
\tilde {\mathbf F}_n).
}
Similar to previous section,
the generalized energy eigenvalues for the discrete state can be
obtained from the $\det \tilde {\mathbf M}(E)=0$, and for each
eigenvalue $E_i$, $\psi_{n\rho}^{(i)}$ can be solved from $\tilde
{\mathbf M}\cdot \tilde {\mathbf Y}=0$. Then
we have
the discrete eigenstates,
\aln{
|\Psi^{(i)}(E_i)\rangle=&\sum_{n'=1}^C\psi^{(i)}_{n'\rho}(E_i)\Big[\sum_{j=1}^D
\frac{f^*_{jn'\rho} }{E_i-M_j}
|j\rangle+{ \sum_{n=1}^C}V_{n\delta',n'\rho}(E_i)\int_{a_n}d\omega\frac{\tilde
g_{n\delta'}(\omega)}{E_i-\omega}
|\omega;n\rangle\Big],
\\
\mathrm{with }\quad& \tilde {\mathbf Y}^{(i)\dagger}(E_i)\cdot  \mathbf V(E_i)\cdot
\tilde{\mathbf W}'(E_i)\cdot \mathbf V(E_i)\cdot  \tilde {\mathbf Y}^{(i)}(E_i)=1
\nonumber
}
with integral contour deformed for resonances and virtual states as before.

\bibliographystyle{apsrev4-1}
\bibliography{Ref}

\end{document}